\documentclass[article,3p]{elsarticle}
\journal{Nuclear Physics B}

\usepackage{amsmath,amssymb}
\usepackage{graphicx}
\usepackage[caption=false]{subfig}
\usepackage{xcolor}
\usepackage{bbold} 
\usepackage{xspace} 
\usepackage{grffile}
\usepackage{slashed}
\usepackage{tikz}
\usetikzlibrary{plotmarks}
\usepackage{rotating,multirow} 
\usepackage[normalem]{ulem}

\newcommand\bs\boldsymbol

\begin{document}

\begin{frontmatter}

\title{Machine Learning as a universal tool for quantitative investigations\\ of phase transitions}

\author{Cinzia Giannetti}
\address{College of Engineering, Swansea University (Bay Campus), Swansea SA1 8EN, UK}
\ead{c.giannetti@swansea.ac.uk}

\author{Biagio Lucini}
\address{Department of Mathematics, Computational Foundry, Swansea University (Bay Campus), Swansea SA1 8EN, UK}
\ead{b.lucini@swansea.ac.uk}

\author{Davide Vadacchino}
\address{INFN, Sezione di Pisa, Largo Pontecorvo 3, I-56127 Pisa, Italy}
\ead{davide.vadacchino@pi.infn.it}

\begin{abstract}
The problem of identifying the phase of a given system for a certain
value of the temperature can be reformulated as a classification
problem in Machine Learning. Taking as a prototype the Ising model and
using the Support Vector Machine as a tool to classify Monte Carlo
generated configurations, we show that the critical region of the
system can be clearly identified and the symmetry that drives the transition can be
reconstructed from the performance of the learning process. The role
of the discrete symmetry of the system in obtaining this result is
discussed. A finite size analysis of the learned Support Vector Machine
decision function allows us to determine the critical temperature and
critical exponents with a precision that is comparable to that of the
most efficient numerical approaches relying on a known Hamiltonian
description of the system. For the determination of the critical
temperature and of the critical exponent connected with the divergence
of the correlation length, other than the availability of a range of
temperatures having information on both phases, the method we propose
does not rest on any physical input on the system, and in
particular is agnostic to its Hamiltonian, its symmetry properties and
its order parameter. Hence, our investigation provides a first significant
step in the direction of devising  robust tools for quantitative
analyses of phase transitions in cases in which an order parameter is
not known.    
\end{abstract}


\end{frontmatter}


\section{Introduction}
\label{sect:introduction}

Phase transitions (a basic overview of which can be found
e.g. in~\cite{Blundell}) are ubiquitous phenomena in Statistical
Mechanics, Condensed Matter and Particle Physics systems. In addition,
applications of the physical concepts related to phase transitions
have been proved successful in investigating problems in other
scientific domains such as the boolean satisfiability problem in
Mathematics (which is an archetypal example of an NP-complete problem,
see e.g.~\cite{HSS}) and cancer dynamics. Some applications beyond
traditional physical systems are discussed for instance in~\cite{Sole}.  

We say that we are in the presence of a phase transition when there is a
point or an hypersurface in parameter space that separates two regions
of the system with very different properties (e.g. the density of ice
is significantly different from that of water, and this change happens
at the freezing point). Mathematically, a phase transition is a
singularity in physical observables as the number of degrees of
freedom of the system goes to infinity. Understanding the dynamics of
the two phases near the transition point and being able to quantify
the location of the latter (in addition to establishing the presence of a
transition, a question that sometimes has not an immediate answer) are
wide reaching issues that have been 
investigated from various angles and perspectives since the early days
of thermal physics, with invaluable insights that have originated some
of the most remarkable ideas in theoretical physics. Two related
examples of transformative ideas originating from the investigation of
phase transitions are the concept of the renormalisation group and the
deep connection between the concept of criticality in Statistical
Mechanics and renormalisability of gauge theories~\cite{Wilson:1973jj}.

The current standard approach to phase transitions relies on a first
principle knowledge of the system. Generally, one investigates a system whose classical or
quantum dynamics is in principle known and can be worked out from an
explicit Hamiltonian or a Lagrangian, respectively. The Hamiltonian
(Lagrangian) has some manifest symmetry that is spontaneously broken
as a function of some control parameters. Based on this, one builds an
order parameter, i.e. an observable that is not invariant under the
relevant symmetry of the Hamiltonian (Lagrangian). In the phase in
which the symmetry is implemented {\em \`a la Wigner}, the lack of
invariance of the order parameter forces this observable to be
zero. Conversely, the fact that the order parameter
observable is different from zero in a phase is an
explicit signal that in that phase the symmetry is not linearly
realised, i.e. one is in the presence of a spontaneously
broken symmetry. A system in which the symmetry is spontaneously
broken possesses a set of degenerate groundstates that transform into
each other under the relevant symmetry group rather than a single,
symmetric groundstate, as it is the case in the symmetric phase.

This {\em ab initio} approach, which is by now consolidated and described
in detail in various textbooks (e.g.~\cite{cardy1996scaling}), has
produced widely accurate results in a variety of contexts, including
Monte Carlo simulations of gauge theories (see~\cite{Langfeld:2015fua} for a recent example). However, there
are relevant physical systems for which the order parameter is hard to
identify and currently unknown, mainly because the symmetry that
drives the phase transition is not manifest in the
Hamiltonian. Remarkable examples in this class are topological phases~\cite{Continentino},
in which the phase transition, being of a topological nature, is
driven by a dual order parameter that might not be immediate to
express in terms of the local variables or indeed might not be known
in terms of the latter, and Quantum Chromodynamics (the theory of the
strong force) at finite quark mass, for which it is still debated what
is the mechanism that drives the phase transition (when it exists, as
a function of the constituent quark mass) and whether it is possibly
of a topological nature (e.g.~\cite{DElia:2005sfk}.)

Recently, a surge of interest has been generated by the possibility of
using Machine Learning inspired techniques for identifying phase
transitions~\cite{Carrasquilla2017}. The underlying idea is to
use clustering techniques to identify properties of the phase
transitions without any {\em a priori} information (a setup that in
the Machine Learning terminology is referred to as {\em unsupervised
  learning}) or by using particular realisations of the system for
which the phase is known unambiguously to understand whether there is a
critical set of parameters for which the phase transition takes place
({\em supervised learning}).~Both supervised and unsupervised learning
characterisations of phase transitions have produced encouraging first
results for identifying and studying phase transitions in Condensed
Matter, Statistical Mechanics and Quantum Field Theory. An incomplete
set of references is provided by~\cite{Carrasquilla2017,Wang,
  Tanaka:2016rtu, Nieuwenburg, Ponte, trebst, Wetzel:2017ooo,
  PhysRevE.96.022140, PhysRevE.95.062122, 2017arXiv170804622M,
  PhysRevB.97.045207, 2018arXiv180909927R, Iso:2018yqu,
  Kashiwa:2018jdi, Cossu:2018pxj,Ohtsuki, Mano, PhysRevB.95.245134,
  Broecker, Chng, Zhang, Zhang:2017iji, PhysRevB.97.205110,
  doi:10.7566/JPSJ.87.033001,PhysRevB99SVM,PhysRevB99SVM2},  
with a recent review given in~\cite{Carrasquilla2017-2}.      

To our knowledge, so far all studies have focused on qualitative and
semi-quantitative results using a varying degree of {\em a priori}
knowledge on the target system. In this paper, we shall investigate
whether it is possible to identify the critical region and
characterise it from a quantitative point of 
view by using Machine Learning with a minimal number of assumptions. To
be more specific, we will ask whether from the simple knowledge of
states of the system at various temperatures we can predict whether a
phase transition takes place and in case extract
precise values of observables and critical quantities such as the
critical temperature and critical exponents as the system undergoes the
transition. As the system of choice for this analysis we have taken
the Ising model in two dimensions, which has an exact analytical
solution and can be investigated numerically with efficient Monte
Carlo techniques. From the Machine Learning point of view, we do the
investigation with a Support Vector Machine (SVM) analysis of Monte
Carlo generated data. This will be contrasted to a traditional
analysis of the same Monte Carlo data. One of the characteristics of
the SVM that makes it particularly suitable for investigating
quantitative issues is that its predictions are based on controlled
analytical models whose parameters are extracted with well-defined
optimisation procedures. Once the model reconstructed by the algorithm
on the data is known, in principle one can use it to get insights on
the physical phenomenon that drives the transition. Among recent studies of phase transitions
with Machine Learning techniques, our work share a similar approach
with~\cite{Ponte}.  However, there are significant differences between
our investigation and the latter reference (for instance, the training
strategy), on which we shall return later.  The main findings
of our study are: (a) for the Ising model, the critical region is
easily identified by training the SVM with two ensembles of 200
configurations, each obtained at temperature values that are
respectively one deep in the ordered phase, the other deep in the disordered phase;
(b) information on the symmetry and on the optimal training
temperatures can be obtained by optimising the performance of the
learned model, the physical case corresponding to the best performance;
(c) once the algorithm has been optimised, a finite size scaling
analysis of the classification function (called the {\em decision
  function} for the SVM) yields results for the critical temperature
and for the critical exponents that are comparable in precision to
those obtained from finite size scaling of an {\em a priori} known
order parameter, even if we have not informed the process with any
previous knowledge on the underlying physics driving the phase transition. 

The rest of the paper is organised as follows. The
formulation of the Ising model and the description of its critical
properties are the subject of Sect.~\ref{sect:ising}, where we also discuss
the Monte Carlo method for generating the data that we have
processed with the SVM. In Sect.~\ref{sect:svg}
we review the mathematical framework underpinning the SVM, with
emphasis on the aspects that have been used in our work. Our numerical analysis using the SVM will be
reported in detail in Sect.~\ref{sect:results}. Finally, our findings are summarised in
Sect.~\ref{sect:conclusions}, where we also outline potential future directions.

\section{The phase transition of the Ising model}
\label{sect:ising}

Due to the simplicity of its formulation coupled to the non-trivial
features it displays, the Ising model is commonly used to illustrate
key concepts and test new techniques in Statistical Mechanics. Its
most direct physical counterpart is a ferromagnet in the vicinity of
the Curie point, but the model can also be reformulated to describe a lattice
gas or a binary alloy. More in general, the Ising model describes an
order-disorder phase transition in dimension two and above. In two
dimensions, it can be
solved analytically in a physically relevant region of parameter
space. It is both the presence of a
non-trivial phase structure and the availability of an analytical
solution that make the Ising model in two dimensions an ideal test bed for
new approaches and ideas in Statistical Mechanics, Condensed Matter
and Lattice Field Theory. In this work, we shall use the model to
explore whether a quantitative study of the phase transition (more
specifically, critical temperature and critical exponents) can be
performed using Machine Learning methods. In order to provide a
comparison of these techniques with a more traditional Monte
Carlo approach, in this section we present a study with the latter
method.  

The Ising model is defined  by the Hamiltonian
\begin{equation}
	\mathcal{H} = -\mathcal{J}\sum_{\langle i, j \rangle} \sigma_i \sigma_j - h \sum_i \sigma_i~,
\end{equation}
with the $\sigma$'s, which take the values $\pm 1$, being conventionally referred as spins. Each
spin is defined on the sites of a two dimensional lattice that we take
to be a squared grid with equal spacings in the two orthogonal directions and closed with periodic
boundary conditions. Each side of the lattice
has total length $L$ and the total area occupied by the system is given by $V =
L^2$. $\sum_{\langle i,j\rangle}$ indicates a sum over nearest
neighbours. $\mathcal{J}$ is the nearest-neighbour spin-spin
coupling (that we choose ferromagnetic, i.e. $\mathcal{J}>0$).  $h$ is an
externally applied field (in the traditional ferromagnetic language,
which we follow from  now on, $h$ is an external magnetic field),
coupled linearly with each spin. 

At vanishing external magnetic field, the state of lowest (internal) energy is easily seen to be an
\emph{ordered} state, in which all of the $\sigma$'s have the same
value, $+1$ or $-1$.  A non-zero value of $h$ splits the degeneracy of
those two states, with the ground state having all spins aligned to $h$. At finite
temperature $T = 1/ (k \beta)$, with $k$ the Boltzmann constant, the
probability of finding the system in a configuration with spins taking
the values $\{\sigma_i\}$ is given by 
\begin{equation}\label{eq:boltz_ising}
  p(\{\sigma_i\}) = \frac{1}{Z}~e^{-\beta \mathcal{H}} \ ,
\end{equation}
where $Z$ is the partition function
\begin{equation}
	Z(\beta, h) = \sum_{\left\{ \sigma_i =\pm 1\right\}} e^{-\beta
          \mathcal{H}} =  e^{- \beta F} \ , 
\end{equation}
with $F$ the free energy of the system. The sum defining $Z$ is taken
over all possible values of the spins. For later use, we define the
ensemble average of an observable $O$ depending on the spin variables
$\sigma_i$ as
\begin{eqnarray}
\langle O \left( \left\{ \sigma_i \right\} \right)\rangle  = 
\frac{1}{Z} \sum_{\left\{ \sigma_i =\pm 1\right\}} O \left( \left\{ \sigma_i \right\} \right) e^{-\beta \mathcal{H}} \ .
\end{eqnarray}
Let us now consider for simplicity the case $h = 0$. Due to the fact that the weights $e^{-\beta\mathcal{H}}/Z $ are
positive definite for any realisation of the spin configuration
$\left\{ \sigma_i =\pm 1\right\}$, the expression defining $O$ has a simple
interpretation as the average over the probability distribution
provided by the normalised Boltzmann factor
\begin{eqnarray}
\label{eq:weight}
P(\mathcal{H} = E) = \rho(E) e^{- \beta E}/Z \ , 
\end{eqnarray} 
where the function $\rho(E)$, which counts the number of
configurations giving $\mathcal{H} = E$, is known as the {\em density
of states}. In terms of $\rho(E)$ we can rewrite 
\begin{equation}
\label{eq:partitionfunction}  
  Z(\beta) = \sum_{E} \rho(E) e^{-\beta E}  \ . 
\end{equation}

For low temperatures (corresponding to large $\beta$), we expect the Ising
system to have a significant number of spins pointing to the same
direction. While this direction is arbitrary, the dynamics will force
the system to choose one, and tunnelling between the two will be
exponentially suppressed with the size of the system. The dynamical
selection of a particular state, out of a set of symmetrically connected ones
with the same energy, realises the key concept of {\em spontaneous
  symmetry breaking}. In our model, the Hamiltonian $\mathcal{H}$ is
invariant under the simultaneous transformation of all the spins
\begin{equation}
\sigma_i \mapsto - \sigma_i \ ,
\end{equation}
which is implemented by the (global) symmetry group $\mathbb{Z}_2
\equiv \{-1,1\}$.  A $\mathbb{Z}_2$ transformation, however, does not leave invariant the two degenerate
groundstates, but interchanges them. For this reason, the choice of a
groundstate (or more in general of a preferred direction of spin
alignment) over the other breaks the invariance of the system. The
expression {\em spontaneous symmetry breaking} underlines the crucial
fact that the global symmetry of the Hamiltonian is broken by the dynamics rather
than by some explicit coupling. 

In the opposite limit of very high temperature,
the energy component of the free energy of the system becomes
negligible if compared to the entropy term. In this regime, spins are effectively randomised, with no
clear alignment being visible. In this phase, the $\mathbb{Z}_2$
symmetry of the system is restored. The phase with spontaneously
broken symmetry is separated from the symmetric phase by the
critical value of the temperature $T_c$ given by
\begin{equation}
k T_c = 2 {\cal J}/ \left( k \log \left( 1 + \sqrt{2} \right) \right)
\ .
\end{equation}
In general, a quantity that allows us to distinguish in which phase we are
is called an {\em order parameter}. The order parameter of the Ising
model is the \emph{reduced magnetisation} $m = M/V = \langle \sum_i \sigma_i/V
\rangle$ ($M = \langle \sum_i \sigma_i \rangle$ is called the {\em
  total magnetisation}) . For this observable, in the limit $L \to \infty$, one finds
\begin{equation}
	|m| = \big| \langle \sum_i \sigma_i \rangle \big|/V = 
        \begin{cases} =0 &\mbox{$\beta <\beta_c$,}\\
		\neq 0 &\mbox{$\beta > \beta_c$~.}
	\end{cases}
\end{equation}

Another important quantity is the correlation length 
$\xi$, which can be understood as the range of the
effective interactions or, equivalently, the typical size of a region ({\em
  cluster}) over which spins are aligned. $\xi$ is formally defined
from the expected scaling of the correlation of two spins $\sigma_l$
and $\sigma_m$ sitting at points $l$ and $m$ as 
\begin{eqnarray}
\langle \sigma_l \sigma_m \rangle - \langle \sigma_l
  \rangle\langle \sigma_m\rangle \ \mathop{\sim}^{r \to \infty} \ \frac{e^{-r /\xi}}{r^p} \ , 
\end{eqnarray}
where $r$ is the distance between $l$ and $m$. $p$ is a calculable
exponent that near the critical temperature is given by
\begin{equation}
p = d - 2 + \eta \ , 
\end{equation}
where $d$ is the dimension of the system ($d = 2$ in our case) and
$\eta$ is a dynamical exponent called {\em anomalous dimension}. 

As the critical point is approached, clusters grow in size and,
exactly at the critical temperature, clusters of all sizes are
present. At this point, the system is invariant with respect to a
scale transformation and the correlation length is infinite. The divergence
of $\xi$ as $t \to 0$ is observed to behave as a power law, i.e. 
\begin{equation}
	\label{eq:crit_nu}
	\xi\propto |t|^{-\nu}~,
\end{equation}
where $t=(T_c-T)/T_c$ is called the {\em reduced temperature} and $\nu$ is
the \emph{thermal} critical exponent.

Other thermodynamic quantities sensitive to the phase transition like
the magnetisation $m$ and the magnetic susceptibility
$\chi = V \left( \langle \sum_{i,j} \sigma_i \sigma_j \rangle/V^2 - m^2 \right)$  have power-low
singularities as $T_c$ is approached:
\begin{equation}
  |m| \mathop{\propto}_{t \to 0^+} t^{\beta} \ ,\qquad \chi
  \mathop{\propto}_{t \to 0}  t^{- \gamma} \ , 
\end{equation}
where $\beta$ and $\gamma$ are two additional (calculable in our case)
critical exponents. 

The final two phenomenologically relevant critical exponents, $\delta$
and $\alpha$, are defined from the following behaviours:
\begin{equation}
\big| m \big| _{t = 0} \propto |h|^{1/\delta} \ ;
\end{equation}
\begin{equation}
c_V= \frac{1}{V} \left( \langle  H^2 \rangle - \langle H \rangle^2 \right)
  \propto t^{- \alpha} 
\ ,
\end{equation}  
$c_V$ being the specific heat of the system. 

The power law behaviour of the above or similarly defined quantities (and hence the
existence of critical exponents) is a general feature of second order
phase transitions, and not only a characteristic of the Ising model. 
Hence, an essential aspect of any study (numerical or analytical) of a
phase transition is the derivation of its critical exponents. The
importance of these quantities is highlighted by the phenomenon of
{\em universality}: systems with very different interactions, but with the
same symmetry structure and having the same dimensionality, share the
same critical behaviour.  Therefore, rather  than being specific to
the model, the set of critical exponents  $(\alpha, \beta, \gamma,
\delta, \nu,\eta)$ only depend on the dimensionality of the system and
on the symmetry of its Hamiltonian, and not on the microscopic
structure of the latter. Universality makes the Ising model very
relevant for studies of systems with more complicated 
Hamiltonians that display the same global $\mathbb{Z}_2$ symmetry. 

Using the scaling hypothesis, which assumes that the free energy is an
homogeneous function of $t$ and $h$ with respect to rescaling of
lengths by an arbitrary factor of $b$, one can derive the following scaling
relations:
\begin{align}
\begin{array}{l l}
\text{Fisher Law:} & \gamma = \nu(2-\eta) \ , \\ 
\text{Widom Law:} & \gamma = \beta(\delta -1) \ ,\\
\text{Rushbrooke Law:} & \alpha + 2\beta + \gamma = 2 \ ,\\
\text{Josephson Law:} & \nu d = 2- \alpha \ ,
\end{array}
\end{align}
which show that only two of the critical exponents are independent,
while the others can be derived. The two critical exponents that are
directly related to the rescaling of $t$ and $h$ are respectively
$\nu$ and $\eta$. Hence, based on this physical motivation, these two
exponents are considered fundamental, while the others are considered secondary.

While in the specific case of the two-dimensional Ising model the
availability of an explicit solution allows
us to compute the critical exponents and the critical temperature, in
more general settings one has to resort to first-principle numerical
techniques. In a Monte Carlo based numerical approach,  two aspects
would need to be considered: (i) the generation of a
sample of configurations of the system according to the Boltzmann
weight Eq.~(\ref{eq:weight}), in order to compute observables in a
controlled way; and (ii) the extraction of
critical exponents from the scaling of key observables with the size
of the system. 

For the first step above, a Markovian process is defined that allows one to obtain
a chain of configurations distributed according to the Boltzmann
weight computed at the target temperature. The physics of the
system plays a crucial role in designing an efficient Markov
process. In the case of the Ising model, the Wolff
algorithm~\cite{Wolffalg} provides the most suitable method for exploring
the configuration space. According to this algorithm, in order to
generate a configuration from a preceding one, we
start from a randomly chosen spin, called \emph{seed}, from which a
cluster is grown by adding to it neighbour spins with the same
orientation with probability $1-e^{-2 \beta \mathcal{J}}$. The growth process
proceeds by exploring the neighbourhood of the newly added spins and
applying the same growth rule until all equally oriented spins connected to the seed have
been proposed for addition, at which point the whole cluster is
flipped. This process defines a new configuration that becomes the
next element of the Markov chain. If one starts from a random
configuration, general principles of Markov processes guarantee the
convergence to the equilibrium distribution given
by~(\ref{eq:weight}). With a chain of $N$ thermalised configurations $C_j$,
the thermal average of an observable $O$ can be written as
\begin{equation}
\langle O \rangle \simeq \sum_{j} O(C_j) e^{- \beta H(C_j)}/
\left( \sum_j e^{- \beta H(C_j)} \right) \ , 
\end{equation}
where the convergence of the approximated value to the exact one is
${\cal O}(1/\sqrt{N})$. Since the latter is a statistical controllable
error, the method is first principles, in the sense that there is a
rigorous way for approximating the exact result at any specified level
of precision. While this is a general fact, we remark for
  completeness that in practical applications of Monte Carlo methods
  systematic errors may arise. The most common ones are due to
  thermalisation (i.e. if not enough configurations are discarded
  before the Markov process reaches the stationary distribution),
  autocorrelation (characterised as lack of enough information in the
  sample due to slow dynamics of the Markov process) and lack of
  ergodic exploration of the configuration space (due, e.g., to the
  presence of disconnected topological sectors). The systematics is well
  under control in the two-dimensional Ising model, thanks to the
  availability of efficient algorithms that have been tested against
  the known solution. Powerful general tests also exist for systems
  where an analytic solution is not known, although these tests (often
  based on comparisons with semi-analytic or perturbative approaches) are
  only as good as our prior knowledge of the broad physical properties
  of the system.\footnote{For instance, if we do not know about the
    existence of specific topological sectors, we will not be able to
    test for the ergodic exploration of the latters.}

In general, the values of the critical exponents can be obtained from
the use of \emph{finite size scaling}, whereby the size of the system
$L$ enters scale-invariance arguments as a renormalisation group {\em
  relevant} quantity with length dimension one. Given that one can build the adimensional
ratio $\xi/L$, we can trade the correlation length $\xi$ with $L$,
which does not introduce new scaling exponents. Under the assumption
that the finite volume corrections to scaling are analytic in $\xi/L$,
one can derive the textbook relation 
\begin{equation}
  \label{eq:critbetac}
	\beta_c - \beta_c(L) \propto L^{-1/\nu} \ , 
\end{equation}
where\footnote{For simplicity, from now on we set $k = 1$.} $\beta_c(L) = (T_c(L))^{-1}$ is the value
at which a susceptibility such as $\chi$ computed at volume $V =L^2$
achieves its maximum.\footnote{$T_c(L)$ is called the {\em
  pseudocritical temperature}.} This maximum value, which will be referred to as
$\chi_{\text{max}}$, diverges as $L \to \infty$, with its position
converging to $\beta_c$. Using scaling arguments, one can show that  
\begin{equation}\label{eq:critgammanu}
	\chi_{\text{max}} \propto L^{\gamma/\nu} \ .
\end{equation}
The whole set of critical exponents can be reconstructed by measuring the value of
$\chi_{\text{max}}(L)$ and of its position $\beta_c(L)$ and using the
asymptotic behaviours provided in Eqs.~(\ref{eq:critbetac},\ref{eq:critgammanu}) to
determine $\nu$ and $\gamma$. In practical applications,
  these exponents are extracted through a fit on a set of data
  obtained at different volumes. In this process, one has to consider that these
  arguments are asymptotic. A systematic error can henceforth arise if
  the volumes explored are not in the asymptotic region. This error is
  generally controlled by repeating the analysis discarding smaller volumes and adding larger
  ones, until a regime of convergence is determined. In the Ising
  case, one can cross-check the results against the analytic
  solutions. When an analytic solution is not available, comparisons
  with other approaches (e.g. predictions in a $4 - \epsilon$
  expansion, when the latter is sufficiently reliable) can be
  instructive. Another source of systematic errors come from scaling
  violations, which, however, general arguments show to be sub-leading
  and undetectable at the level of precision that can be reached in
  standard simulations, their identification requiring dedicated
  methodologies. Hence, we neglect them for the remainder of our discussion.

Using conventional analysis of Monte Carlo simulated data  based on
scaling arguments, we extracted numerically the critical temperature
and the critical exponents of the model, comparing our determination
to the known exact results. We stress that this approach is very well
established and has been used for the study of phase transitions in
various Condensed Matter, Statistical Mechanics and Field Theory models
(including the Ising model) for a long time. The reason why we
repropose it here is to be able to assess the quality of our SVM analysis,
comparing the results on the same set of input data. 

Our Monte Carlo simulations were run on $L^2$ lattices of linear size
ranging from $L=32$ to $L=1024$ and for several values of
temperatures between $T=0.5$ and $T=5.0$. Near the transition, $3000$
Wolff clusters were flipped for thermalisation, in order to allow the
system to relax to its equilibrium state, and the configurations
were recorded once every $15$ Wolff updates.  Far from the transition,
the separation between configurations was reduced, as temporal
correlations of the Markov chain are less severe. For each lattice size
and temperature value, we recorded $200$ configurations.  For the scaling analysis that allowed us
to extract $\beta_c$, $\nu$ and $\gamma$ we used data for order $20$
equally spaced temperature values $T$ in the critical region, i.e. in a small
neighbourhood of $T_c$ in which one can verify a posteriori that
scaling arguments apply. The simulated values are reported in Tab.~\ref{tab:betac}.

\begin{table}

\begin{center}
\begin{tabular}{cccc}
\hline
$L$ & $T_\text{min}$ & $T_\text{max}$ & $n_\text{steps}$ \\
\hline
$64  $ & $2.280$ & $2.330$ & $20$ \\
$128 $ & $2.275$ & $2.294$ & $20$ \\
$240 $ & $2.273$ & $2.285$ & $24$ \\
$360 $ & $2.270$ & $2.280$ & $20$ \\
$440 $ & $2.270$ & $2.280$ & $20$ \\
$512 $ & $2.2665$ & $2.2770$ & $22$ \\
$760 $ & $2.27000$ & $2.27400$ & $20$ \\
$1024$ & $2.27000$ & $2.27300$ & $30$ \\
\hline
\end{tabular}
\qquad
\begin{tabular}{ccc}
\hline
$L$ & $T_c$ & $\chi_\text{max}$ \\
\hline
$ 64$ & $ 2.3037(29) $ & $  1.284(37) \cdot 10^2$ \\
$ 128 $ & $ 2.28664(74) $ & $  4.590(97) \cdot 10^2$ \\
$ 360 $ & $ 2.27528(28) $ & $  2.781(65) \cdot 10^2$ \\
$ 440 $ & $ 2.27448(47) $ & $  3.97(10)\cdot 10^3 $ \\
$ 512 $ & $ 2.27351(29) $ & $  5.24(14)\cdot 10^3 $ \\
$ 240 $ & $ 2.27892(39) $ & $  1.383(28) \cdot 10^3 $ \\
$ 760 $ & $ 2.27226(25) $ & $  1.035(21)\cdot 10^4 $ \\
$ 1024 $ & $ 2.27145(23) $ & $  1.757(40)\cdot 10^4 $ \\
\hline
\end{tabular}
\end{center}
\caption{On the left, scanning windows of temperatures for extracting
  the pseudocritical temperature $T_c(L)$ at each value of $L$; $n_\text{steps}$ indicates
  the number of simulated values of $T$, all equally spaced between
  the two extremes $T_\text{min}$ and $T_\text{max}$. On the right,
  values of the pseudocritical temperature $T_c$ and the corresponding
  maximum of the magnetic susceptibility, as obtained from the
  multi-histogram method.} 
\label{tab:betac}
\end{table}

\begin{figure}[th]
  \subfloat{\includegraphics[scale=1.]{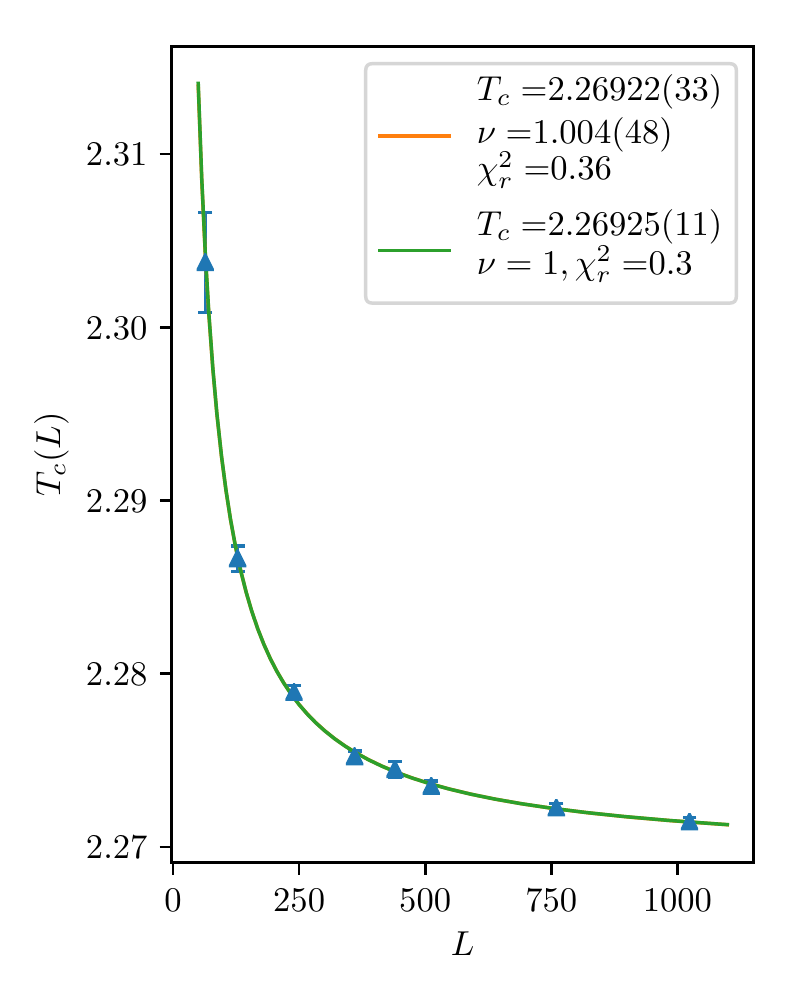}}
  \subfloat{\includegraphics[scale=1.]{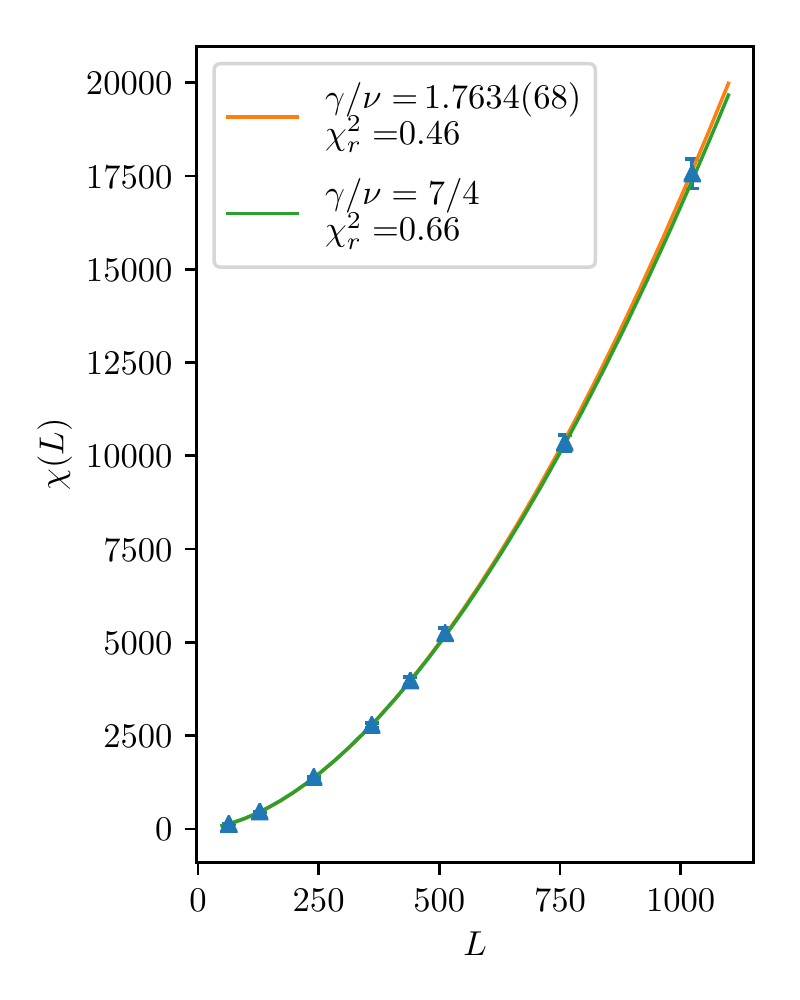}}
\caption{(Colour online) Left: behaviour of the pseudocritical temperature with the
  linear size of the lattice. Right: behaviour of the peak value of
  magnetic susceptibility with the linear size of the lattice. A fit
  to the expected asymptotic functional form is also displayed. 
\label{fig:plots}}
\end{figure}

To extract the infinite volume critical coupling and the critical
exponents $\nu$ and $\gamma/\nu$, we fitted Eq.~(\ref{eq:critbetac})
and Eq.~(\ref{eq:critgammanu}) to the data in the critical region
using $\beta_c(\infty) \equiv \beta_c$,
$\nu$ and $\gamma/\nu$ as fitting parameters after reweighting the
measured observables with the multi-histogram method (see
Appendix~\ref{sect:app:reweighting}). Errors were estimated using
bootstrap (described in Appendix~\ref{sect:app:bootstrap}). The results
can be found in the first row of Tab.~\ref{tab:fits}.\footnote{From
  here onwards, we set ${\cal J} = 1$.} In the same
table, the second row reports fits in which the values of the
critical exponents are fixed to their analytically computed
values. The fit results are visible in Fig.~\ref{fig:plots}.  

As expected, the method reproduces well the analytically
computable results,  with a precision of order $10^{-4}$ on $T_c$,
and of $10^{-3}$ on $\gamma/\nu$ (with the central value in this latter case being
compatible with the expected value within two standard deviations,
while in the former the compatibility is within the statistical error).
The critical exponent $\nu$ is determined with a precision of the order of five
percent. More precise results can be obtained by increasing the
number of generated configurations and/or the set of
simulated temperatures, which in this model can be
achieved with a moderate increase in computational time. However, we stress that the purpose
of the study we have discussed is to establish a numerical
benchmark for the SVM analysis provided in Sect.~\ref{sect:results} using the
same input information, rather than performing a high precision
investigation of the phase transition in the Ising model with finite
size scaling techniques, which is by now a classic topic in
specialised textbooks (see e.g.~\cite{landau2014guide}).   

\begin{table}
\begin{center}
	\begin{tabular}{ccc|cc}
		\hline
		$T_c$ & $\nu$ & $\chi^2_r$ & $\gamma/\nu$ & $\chi^2_r$\\
		\hline
		$2.26922(33)$ & $1.004(48)$ & $0.36$ & $1.7634(68)$ & $0.46$ \\
		$2.26925(11)$ & $1$ (exact) & $0.3$ & $7/4$ (exact)&  $0.66$ \\
		\hline
	\end{tabular}
\end{center}
  \caption{On the left, the extracted values of $T_c$ and $\nu$
          obtained from fitting Eq.~(\ref{eq:critbetac}) to the data
          at the simulated values of $L$. On the right, the extracted
          values of $\gamma/\nu$ obtained from fitting
          Eq.~(\ref{eq:critgammanu}) to the data. We also show
          $\chi^2_r$, the $\chi^2$ per degree of freedom, of each
          fit.}  
	\label{tab:fits}
\end{table}

\section{Ensemble classification and the Support Vector Machine}
\label{sect:svg}

The Support Vector Machine (SVM) is a popular supervised learning
algorithm used to solve classification and regression problems. In the
field of Machine Learning, supervised learning approaches use labelled
training data (i.e. data for which the classification is known) to find a
model describing the functional relationship $Y=f(X)$ between a
response variable $Y$ (where most often $Y$ is a label, i.e. a set of
discrete values) and input variable(s) $X$. The learned model
enables one to predict values of $Y$ for previously 
unseen values of $X$. Supervised learning is often contrasted to
unsupervised learning approaches. The goal of unsupervised learning is
to model the underlying structure of the data to discover patterns
(for instance finding clusters of observations) and insightful
representations rather than predicting a functional relationship. Unike in supervised learning, in this case the learning
process uses only the input values $X$, assuming no knowledge of any
output. Both supervised and unsupervised methods are widely used for modelling and
prediction for a variety of applications including fraud detection,
image and speech recognition, quality control and defect or failure
prediction. In addition to SVM, supervised learning techniques include
Artificial Neural Network, Decision Trees and K-Nearest Neighbour,
while unsupervised learning can be achieved through clustering
(k-means or EM Clustering), Kernel Density Estimation, Principal
Component Analysis (PCA) and Self Organising Maps (SOM). 

SVM was firstly introduced in 1963 by Vladimir Vapnik and Alexey
Chervonenkis, and further
developed in the 90's as a general solution to linear binary and
multi-class classification problems, and as the generalisation thereof
to cases in which the target data are not linearly
separable~\cite{cortes1995support, boser1992training}. The SVM method
can also be modified to solve regression problems when the label can
take continuous real values instead of categorical values. The SVM
method is very effective in high dimensional spaces where the number of
dimensions is greater than the number of samples (see
e.g.~\cite{Pappu2014}). It also differs from other supervised learning 
techniques such as Artificial Neural Network because it can be
expressed as a convex optimisation problem and, if a solution exists,
it is always found as the unique global minimum of the equivalent
optimisation problem~\cite{shawe2004kernel}. The simple visual interpretation of the
procedure and the possibility to use a wide range of functional forms (provided by
transformations that in this context are called {\rm kernels}) make
the SVM an effective method
in problems when boundaries that separate the data (the {\em decision
  boundaries}) can not be expressed as a linear hyperplane in the space
of the input variables~\cite{kecman2001learning}. 

\begin{figure}[tbp]
  \subfloat[Example of linearly separable data.]{\includegraphics{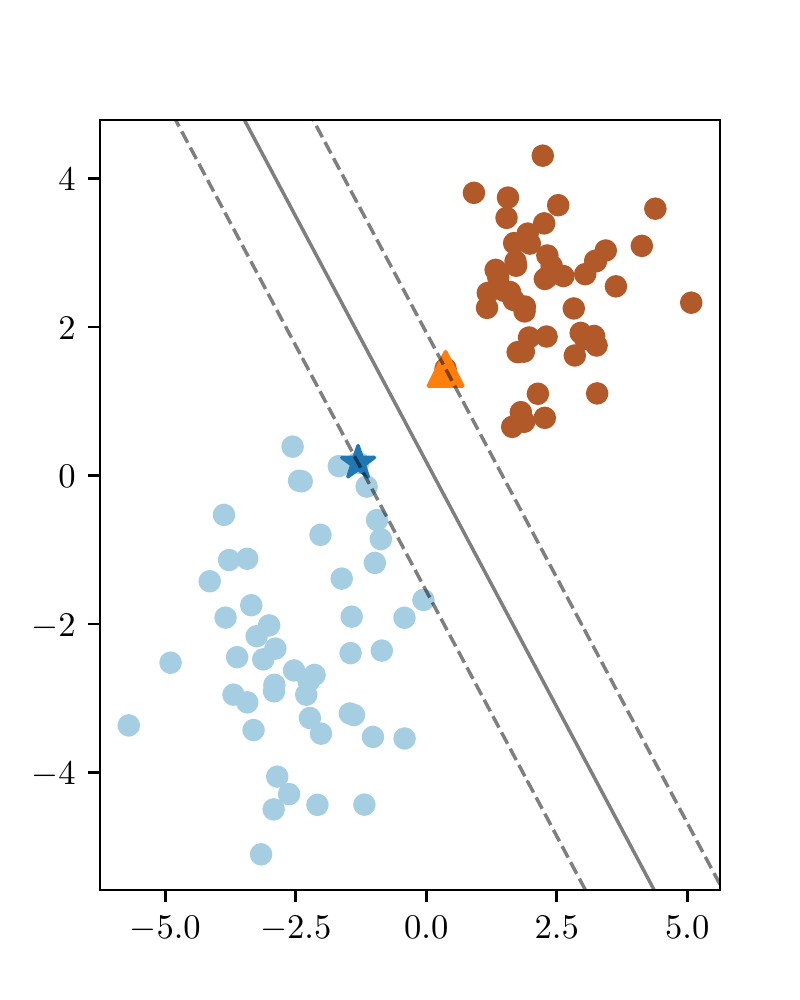}}
  \subfloat[Example of non-linearly separable data.]{\includegraphics{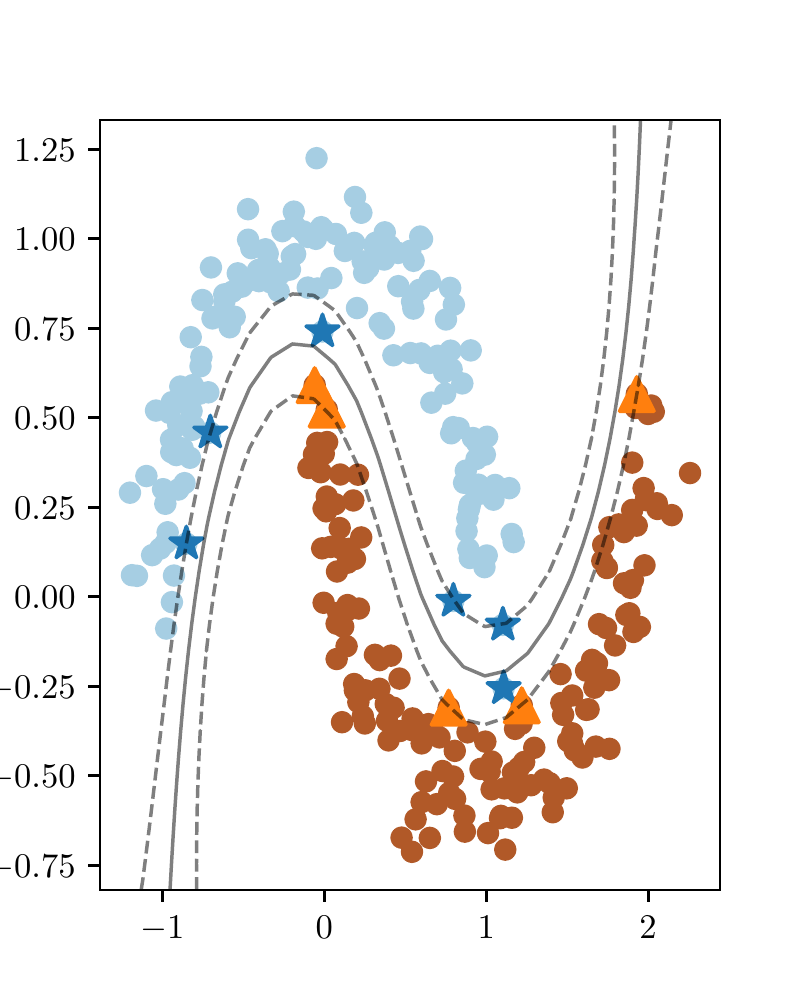}}
  \caption{\label{fig:maxmargin} (Colour online) Two archetypal examples of
    distributions of training data in a two-class SVM classification problem, taken for
    simplicity to be bidimensional. Circles with the same shade
    indicate points in the same known class and each axis represents a
    feature. The separating hyperplane or hypersurface is denoted with a solid line and the
    margin on either side is indicated by a dashed line. The support
    vectors for each of the two classes are represented respectively
    with stars (corresponding to the lighter points) and with
    triangles (corresponding to the darker points).} 
\end{figure}

In the most straightforward case, an SVM deals with a binary classification
problem between linearly separable datasets. For this problem, in the
space of the input variables, the SVM method seeks to find the
maximum margin hyperplane separating data belonging to the two classes. 
Given $N$ linearly separable training points $(x_i,y_i)$, with $x_i
\in \mathbb{R}^d$ and $y_i \in \lbrace-1, 1\rbrace $, there are
generally many possible separating hyperplanes, all specified in
$d$-dimensional space by the equation 
\begin{equation}
  \vec{\omega}\cdot\vec{x}- \hat{\beta}=0,
\end{equation}
for appropriate values of $\vec{\omega}$ and $\hat{\beta}$, with $\vec{\omega}$ the normal
vector to the plane and $\hat{\beta}$ its offset with respect to the origin.   
For each of these hyperplanes, a {\em margin} can be defined as the region of space delimited by
the two hyperplanes that are parallel to the separating hyperplane and pass
through the closest data points that lie on either side of it. These
points are called support vectors. For each support vector, we define 
the size as the distance between that support vector
and the separating hyperplane. As the name suggests, a Support Vector
Machine is an algorithm that seeks to find the maximum margin
hyperplane, as determined by its support vectors. This plane is
identified by the equation
\begin{equation}
  \vec{w}\cdot\vec{x}-b=0 \ , 
\end{equation}
where $\vec{w}$ and $b$ are obtained through a maximisation process as
discussed below. Once $\vec{w}$ and $b$ have been found,
we are still free to rescale them such that 
\begin{equation}
  \vec{w}\cdot\vec{x}-b=1,\quad \vec{w}\cdot\vec{x}-b=-1 
\end{equation}
for the \emph{support vectors}, where we fix the convention that the
first equation holds for support vectors corresponding to label $+1$
and the other for those with label $-1$.  The size of the {\em margin}
(i.e. the maximum separating slab) will be $2/|\vec{w}|$ and for
points $\left\{ \vec{x}_i, ~y_i \right\}$ on either side of the maximal margin hyperplane, 
$y_i(\vec{w}\cdot\vec{x}_i-b)\geq1$. A diagram that illustrates the
linearly separable case is provided in Fig.~\ref{fig:maxmargin},~left.

Finding the maximum margin hyperplane can be reformulated mathematically as the minimisation
problem  
\begin{equation}\label{eq:minmax_hard}
  \min_{\vec{\omega},\hat{\beta}} ||\vec{\omega}||^2\text{
    s.t. }y_i(\vec{\omega}\cdot\vec{x}_i- \hat{\beta})\geq1,~ \forall i=1,\dots,N \ , 
\end{equation}
where $N$ is the total number of training data. The solution of this  problem delivers the values of $\vec{w}$
and $b$ that allow us to define the desired classifier
\begin{equation}
  f(\vec{x}) = \mathrm{sign} \left( \vec{w}\cdot\vec{x} - b
  \right) 
\end{equation}
assigning a value $\pm1$ to any pervasively unseen $\vec{x}$ depending
on whether it lays above or below the maximal separting hyperplane. The \emph{decision function} is defined as
\begin{equation}
  d(\vec{x}) = \vec{w}\cdot\vec{x}-b
\end{equation}
and its (signed) value determines the distance of $\vec{x}$ from the separating hyperplane.

We now proceed to generalise the above picture by relaxing the
assumption of strict linear separability. We introduce the \emph{slack}
variables $\xi_i$ defined as
\begin{equation}
  \xi_i = \max(~0,~1-y_i(\vec{w}\cdot\vec{x}_i-b)~)~.
\end{equation}
If the data are such that $\left\{ \xi_i =0\right\}$, we fall back to
the linearly separable case. When some of the $\xi_i$'s are
non-vanishing, the corresponding data points fall into the margin. In
that case, we associate a penalty $C$ to each of the non-vanishing
$\xi_i$'s and we modify condition~(\ref{eq:minmax_hard}) as 
\begin{eqnarray}
    \min_{\vec{\omega},\hat{\beta}}\left( ||\vec{\omega}||^2 + \frac{C}{N}\sum_{i=1}^N
      \xi_i\right) \ \text{ s.t. } \xi_i=\max(0,~1-
    y_i(\vec{\omega}\cdot\vec{x}_i-\hat{\beta}))~ \forall i~. 
\end{eqnarray}

For $C=0$, the unique solution to the minimisation problem can only be found in
the linearly separable case, as a hard-margin classifier. For non
vanishing $C$, the solution is called a soft-margin classifier. Using
Lagrangian language, the problem is to minimise 
\begin{eqnarray}
L = \frac{1}{2} ||\vec{\omega}||^2 + \frac{C}{N}\sum_{i=1}^N \xi_i + 
\sum_{i=1}^N \alpha_i \left( 1- y_i
  \left(\vec{\omega}\cdot\vec{x}_i-\hat{\beta} \right) - \xi_i \right) - \sum_{i=1}^N \eta_i \xi_i
\end{eqnarray}
with respect to $\vec{\omega},~\xi_i,\hat{\beta}$, where $\alpha_i$ and
$\eta_i$ are Lagrange multipliers. Note that, as requested by the Karush-Kuhn-Tucker
conditions,\footnote{The Karush-Kuhn-Tucker conditions~\cite{karush1939,kuhn1951}
  are necessary conditions that need to be satisfied in optimisation
  problems where inequality constraints are present.}
$\alpha_i,~\eta_i \geq 0$. Setting the gradients of $L$ to zero imposing
\begin{equation}
  \delta_{\vec{\omega}}~ L = 0\ , \qquad \delta_{\hat{\beta}}~ L =0 \ ,
  \qquad \delta_{\vec{\xi}}~L =0 
\end{equation}
yields
\begin{equation}
    \vec{\omega} - \sum_{i=1}^N \alpha_i y_i \vec{x}_i =0 \ , \qquad
    \sum_{i=1}^N \alpha_i y_i =0 \ , \qquad
    \frac{C}{N} -\alpha_i - \eta_i =0 \ . 
\end{equation}
Plugging these into the original Lagrangian, one obtains
\begin{eqnarray}\label{eq:dual_formulation}
  L = -\frac{1}{2} \sum_{i,j=1}^N \alpha_i \alpha_j y_i y_j
  \vec{x}_i\cdot \vec{x}_j + \sum_{i=1}^N \alpha_i \
  \text{s.t. } \sum_{i=1}^N \alpha_i y_i =0\text{ and }0 \leq
  N \alpha_i \leq C \ , 
\end{eqnarray}
where now the extremisation would need to be performed over the
$\alpha_i$. If the unique solution exists, we call $\{a_i\}$ the set of the
$\alpha$'s that maximises $L$. 

Eq.~(\ref{eq:dual_formulation}) expresses
the \emph{dual} formulation of the problem of finding the maximally
separating hyperplane, which is the starting point to generalize
the applicability of the method to cases where linear separation is
not possible. Indeed, note that $L$ now only depends on the inner
product between representative vectors of training samples. In cases
in which the problem does not appear to have a solution in
$\mathbb{R}^d$, we might seek one in another space of larger dimensionality. We can then consider a more general Lagrangian functional
\begin{equation}
  \label{eq:dual_kernel}
  \tilde{L} = \frac{1}{2}\sum_{i,j} y_i y_j \alpha_i \alpha_j
  \mathcal{K}(\vec{x}_i, \vec{x}_j) - \sum_{i=1}^n \alpha_i \ ,
\end{equation}
where $\mathcal{K}(x_i, x_j)$ is referred to as a 
\emph{kernel.}\footnote{The process of separating the data through a
  kernel is often called ``kernel trick''. We will not use this expression
  any further in this paper, as it would be dangerously misleading. Indeed, as we
  will discuss in detail for our model, the selection of the optimal kernel is not a mere
  computational expedient, but {\em a posteriori} provides powerful insights
on the physics of the target system.}
Note that in writing Eq.~(\ref{eq:dual_kernel}) we have multiplied the
natural generalisation of~(\ref{eq:dual_formulation}) by a global
$-1$, which means that, in order to follow the standard convention, we
have rewritten our original maximisation problem into a minimisation
problem. 
As it is transparent from Eq.~(\ref{eq:dual_kernel}), the value of the kernel computed on the input data is all the
information that the minimisation algorithm sees. For this reason, the
kernel is often denoted as an \emph{information bottleneck} in this
kind of analyses. Thus, the choice of kernel is crucial.  Obviously, the kernel must be symmetric in its arguments, reflecting the symmetry under exchange of two training sets. If we require that the minimisation problem above has a unique global solution, then we must also require that $\mathcal{K}$ is positive definite (the optimisation problem is then convex). According to Mercer's theorem, there exists then a map $\Phi:\vec{x}\to\Phi(\vec{x})$, called \emph{feature maps}, such that the kernel can be represented as a dot product in some higher dimensional space, $\mathcal{K}(x_i, x_j)=\Phi(\vec{x}_i) \cdot \Phi(\vec{x}_j)$. Other than the conditions above, our choice of kernel must be guided by our knowledge of the system, performance or computational ease. If a solution is found that minimises $\tilde{L}$ in the image space for a
given $\Phi$, we
define the {\em decision function} as 
\begin{equation}
  \label{eq:dfgeneral}
  d(\vec{x})= \sum_{i=1}^N \alpha_i y_i \Phi(\vec{x}_i)\cdot \Phi(\vec{x})
  +b \ .
\end{equation}
As in the linearly separable case, the sign of the decision function determines the predicted label of each input point.  An illustrative representation of a non-linearly separable
data set in the space of the input data (again, for this purpose
assumed to be two-dimensional) is provided in Fig.~\ref{fig:maxmargin}, right.\\
 The structure of a kernel that
  allows us to separate the training data is closely related to the dynamics of the
  system. Henceforth, the selection of a class of kernels embodies assumptions
  about the data to be analysed. The more these assumptions are
  constraining, the smaller is the class of functions
  among which we can choose our kernel. For a comprehensive and agnostic
  analysis, intuitively, we would be led to choose from a very large class of kernels, in order to reduce any
  \emph{bias} due to any possible assumption. However, when we try and
  infer some physical behaviour from the ability of a kernel to
  separate the training set, this latter approach will
  generally reflect in a larger number of parameters that need to be
  optimised for the given data, with a correspondingly higher variance in the final
  result, notably due to fitting noise. There is therefore a trade-off between
  variance and bias that is central in many discussions in Machine
  Learning.\\ 
A set of kernels that will be important for analysis below is the class of polynomial kernels of degree $n$, 
\begin{equation}\label{eq:poly_kernel}
\mathcal{K}(\vec{x}_i,\,\vec{x}_j) = \left( \frac{\vec{x}_i\cdot\vec{x}_j}{\Gamma} + c_0\right)^n = \sum_{a=0}^n \binom{n}{a} c_0^{n-a} \left(\frac{\vec{x}_i\cdot\vec{x}_j}{\Gamma}\right)^a~,
\end{equation}
where $\Gamma$ and $c_0$ are constants and $\cdot$ is the usual scalar product in the $\vec{x}$ space. If $c_0=0$ these are called \emph{homogeneous}, otherwise \emph{inhomogeneous}. The components of the feature map in the inhomogeneous case are all the monomials of degree up to $n$ built from the products of the components of $\vec{x}_i$ and $\vec{x}_j$, respectively. These form a linear space of dimension $\binom{N+n}{n}$. For example, in the case $n=2$
\begin{equation}
  \Phi^{n=2}(\vec{x}) = (x_1^2,\,\dots,\,x_N^2,\,x_1 x_2,\dots,\,x_1x_N,\,x_2x_3,\dots,\,x_2 x_N,\,c_0 x_1,\dots,\,c_0 x_N)~.
\end{equation}
From the right hand side of Eq.~(\ref{eq:poly_kernel}) it is apparent that the inhomogeneous polynomial kernel of degree $n$ can be obtained as a linear combination of homogeneous kernels of degrees up to $n$. In homogeneous kernels, only monomials of degree exactly $n$ are present in the feature map. 
Owing to the important role of homogeneous polynomial kernels in this work, we denote them as 
\begin{equation}
  \mathcal{K}^{(n)}(\vec{x}_i,\,\vec{x}_j) = \left( \frac{\vec{x}_i\cdot\vec{x}_j}{\Gamma} \right)^n~. 
\end{equation}

An important aspect of the analysis performed in our study is related
to the symmetry
properties of the system. We will restrict to the case of a \emph{global internal symmetry}
with respect to a group $G$, acting on $\vec{x}$ as
\begin{equation}
  g\,\vec{x} = \left( g x_1, \dots, g x_N\right),\quad \forall g\in G~.
\end{equation}
As explained in \cite{Haasdonk2007}, a symmetry of the system can be interpreted as a prior knowledge on
the system itself. Indeed if a system is invariant under a symmetry
group $G$, we would expect that only features that are invariant with respect to the action of
$G$ shown above will play an important role. Restricting ourselves to those
features forces the kernel to be \emph{totally invariant} with respect to the
action of the symmetry group $G$:
\begin{equation}
  \mathcal{K}(\vec{x},\,\vec{y}) = \mathcal{K}(g\vec{x},\,\vec{y})=
  \mathcal{K}(\vec{y},\,g\vec{x}),~\forall g \in G \ ,
\end{equation}
where the last equality is implied by the symmetry of the kernel with respect to the exchange of its arguments.

As is shown in \cite{Schulz-Mirbach:1992:ECI:903028} , a
complete\footnote{We call a set of invariant features complete if every orbit of the group can be distinguished by using only
those features. Note that we are not requesting this set to be
minimal, i.e. we allow it to contain features that can be
expressed in terms of others features.} set of invariant features for a finite group $G$ of
order $|G|$ can be obtained by \emph{projecting} the kernel in
Eq.~(\ref{eq:poly_kernel}) with $d=|G|$ on the group. For a
generic function $f(\vec{x})$, the projection is realised as an average
over all possible transforms of the argument:  
\begin{equation}
  \tilde{f}(\vec{x}) = \frac{1}{|G|} \sum_{g\in G} f( g \vec{x})~.
\end{equation}
Similarly, the projection of the inhomogeneous polynomial kernel above reads,
\begin{equation}
  \mathcal{K}_G(\vec{x},\,\vec{y}) = \frac{1}{|G|^2} \sum_{a=0}^n
  \binom{n}{a} c_0^{n-a} \sum_{g, g'\in G}\left(\frac{g \vec{x}_i\cdot
      g^\prime\vec{x}_j}{\Gamma}\right)^a~.
\end{equation}

The transformation law of a polynomial kernel under a symmetry can be
rephrased in terms of the behaviour of its homogeneous components
under that symmetry. Specifically, if, as a consequence of the invariance of the
physical system we would like to classify, a polynomial
kernel is invariant under the action of a symmetry group $G$, it can
only contain homogeneous terms that are invariant under $G$. For instance, in the case of a
global symmetry under the $\mathbb{Z}_2$ group, an invariant kernel
only contains even powers, as odd powers are projected out in the
averaging. Hence, we would expect that a good classifying kernel
(i.e. a kernel implementing a transformation taking us in a space in
which a separating hyperplane can be found) will only contain even
powers. In the search for a good classifier, our prior knowledge of
the symmetry has restricted possible candidate kernels.  
More formally, under general assumptions, for systems
with a discrete symmetry group, one can prove   that the empirical
risk is minimised by kernels that respect the symmetry of the
system.

In a bottom-up approach to phase transitions with SVM methods, we aim to reconstruct physical properties
(among which, the global symmetry of the Hamiltonian) of the system by identifying an
appropriate kernel that allows us to separate the two input
classes (e.g. known phases at two temperatures). Following our previous argument, if such a kernel exists, it must respect the
symmetries of the system. Hence, by identifying transformations under
which this kernel is invariant, we can identify the symmetries of the system. However, the problem of finding
{\em a} classifying kernel with a systematic search through
a generic space with no a priori knowledge is a rather hopeless
process, as any arbitrary function respecting the properties of a good
kernel can be in principle the answer we are looking for. Henceforth, we shall investigate whether, by
looking at classification properties of a finite set of kernels
(e.g. all the monomials up to some degree $p$), one can infer or at
least restrict the symmetry properties of the system being
studied. The naive expectation is that kernels that respect the
underlying symmetry will classify better than those that don't. Taking
this perspective inevitably leads us to discussing which classification should we trust
more among all those that separate the data. In the
remaining of this section, we will discuss a procedure to quantify the
reliability of a classification, which will allow us to discriminate
between bad and good classifiers and to chose the best among the latters.

We will use a model selection technique strongly inspired by
structural risk minimisation. To perform model selection, which relies on an evaluation of the
robustness of the trained model, we estimate the expected risk. The
probability of test error, or expected risk, of a model $f$ on a data
sample $(X,\,Y)$, where $X$ are the data points, $Y$ their class
label, which are related according to the probability distribution
$P(X,\,Y)$ can be defined as 
\begin{equation}\label{eq:exp_risk}
  R[f] = \int \mathrm{d}P(X,\,Y) \mathbb{1}\left( Y,\,f(X)\right)~.
\end{equation}
where the function $\mathbb{1}$ is the indicator function, which
vanishes when its arguments coincide and is equal to $1$ otherwise. 

If we knew the probability measure $P(X,\,Y)$ relating $X$ to $Y$ we
could compute the expected risk for any model $f$. Intuitively, the
expected risk is expected to be higher for models that \emph{do not}
correctly predict the class labels of the data points. In practice,
however, we do not know $P(X,\,Y)$, and we must evaluate it from the
data. The \emph{empirical} risk can be defined as 
\begin{equation}\label{eq:emp_risk}
  R_\text{emp}[f] = \frac{1}{m} \sum_{i=1}^m \mathbb{1}\left(Y_i,\,f(X_i)\right)~,
\end{equation}
where $m$ is the number of data points.

To evaluate this quantity we use a procedure called
\emph{cross-validation}. In general, this method consists in splitting
the data 
in two parts: one is used to train a model and the other to test it,
for example computing Eq.~(\ref{eq:emp_risk}) several times over
different divisions. The simplest form of cross-validation is
\emph{Leave-one-out} (LOO-CV), in which all but a single data point
are used for training, and the former for testing. Repeating the
computation of the risk for every possible choice of the removed data
point yields 
\begin{equation}\label{eq:LOOe}
  R_\text{LOO}[f] = \frac{1}{m} \sum_{i=1}^m \mathbb{1}\left(Y_i,\,f_{m-1}(X_i)\right)~,
\end{equation}
where $f_{m-1}$ is the model obtained after training on a $m-1$ size
sample, i.e. with $i$-th data point removed from the sample. The
$R_\text{LOO}$ is shown to be an unbiased estimator (see theorem 12.9
in \cite{Scholkopf:2001:LKS:559923}) of the expected risk evaluated on
a sample of size $m-1$,  
\begin{equation}
  \langle R_\text{LOO}[f_m]\rangle = \langle R[f_{m-1}] \rangle \ .
\end{equation}

A very useful bound for $R_\text{LOO}$ can be obtained by observing
that the removal, from the sample, of a data point which is not a
support vector cannot alter $R_\text{LOO}$, because that point would
be well classified anyways. Thus  
\begin{equation}
  \langle R_\text{LOO}[f_{m}]\rangle\leq \frac{\langle n_\text{SV}\rangle}{m}~,
\end{equation}
where $n_{SV}$ is the number of support vectors and the average is
taken on the $LOO$ sample. Therefore the number of support vectors found in the learning process
is related to the performance of the SVM on unseen examples. The
comparison between the number of support vectors obtained in training
various models on the same data will thus provide useful additional
information in performing model selection. 

The are two problems with the LOO estimates described above. The first
is that, from a computational point of view, they are very demanding,
since we have to perform the training procedure for  a number of times equal to the size
of sample. The second problem is that they are quite noisy. There are
in principle many implementations of cross-validation that can
circumvent both problems. In this work, we will use stratified
$10$-fold CV estimates. These consist in dividing the sample in $10$
equal bins, chosen so that the classes are equally represented, and
performing the train and test procedure above with each bin removed
in turn. This is shown to yield an
unbiased~\cite{Kohavi:1995:SCB:1643031.1643047} estimator of
$R_\text{LOO}$. 

Related to model selection is the problem of overfitting. Overfitting
occurs when we have a large number of parameters to fix 
that are not constrained enough by the available data. When this
happens, we say that the model is overfitting the training data. In
this case, typically, the classifier is significantly affected by the
statistical noise of the input sample and will be unable to correctly
predict the classification of new data. Procedures such as the LOO and
the $10$-fold CV allow one to identify whether overfitting has
occurred by returning a low cross-validation score.  

\section{An SVM analysis of the phase transition}
\label{sect:results}

Our aim is to estimate the critical temperature $T_c$ and the critical
exponents of the 2D Ising model using an approach based on the SVM. For this, we
must first select the kernel we will use. Following the discussion in
the previous section, we will restrict to polynomial homogeneous
kernels. Secondly, we have to devise a technique to extract the
critical temperature $T_c$ and critical exponents from the trained
model. Before delving into the details of our strategy, let us specify
some technicalities regarding the application of the theory explained
in section~\ref{sect:svg} to the case of our study.   

To train the SVM, we will use $200$ configurations at temperature
$T_1$, labelled as $y=-1$, and $200$ configurations at temperature
$T_2$, labelled as $y=1$, with both sets obtained from Monte Carlo simulations. While
eventually we would like to identify each label with a phase, for the
moment the label does not carry this meaning: we can simply identify
the two classes with the two \emph{training temperatures} $T_1$ and
$T_2$.  Then, \emph{training a SVM at temperatures $T_1$ and $T_2$}
means running the learning algorithm using the configurations obtained
at temperature $T_1$ as the training set with $y=-1$ and those at
$T_2$ as the training set with $y=1$ (we assume for simplicity $T_1 <
T_2$). Each configuration will consist of a $L^2$ component vector $\vec{x}$,
each component corresponding to the elementary variable defined on a
site of a square $L\times L$ lattice.\footnote{The square lattice will
  be mapped to a linear array in typographical order.} When we refer
to the \emph{configurations at a temperature $T$}, we mean a sample of
$200$ independent configurations obtained from Monte Carlo simulations at
temperature $T$. In order to find a separating hyperplane, we will be
using homogeneous polynomial kernels of degree $n$,  
\begin{equation}\label{eq:choice_gamma}
  \mathcal{K}^{(n)}(\vec{x}_i,\,\vec{x}_j) = \left(
    \frac{\vec{x}_i\cdot\vec{x}_j}{\Gamma} \right)^n~, 
\end{equation}
where $i$ and $j$ label the configurations. To fix $\Gamma$, whose
value is irrelevant for the minimisation algorithm, we require that
for a configuration in which $\vec{x} =(1,\,1,\,\dots,\,1)$,
$\mathcal{K}(\vec{x},\vec{x})=1$. Since $\vec{x}$ has $L^2$ unit valued
components, $\vec{x}\cdot\vec{x}=L^2$. Thus, $\Gamma= L^2$. The
corresponding form of the decision function will be 
\begin{equation}\label{eq:decfun_poly}
  d(\vec{x}) = \frac{1}{L^{2n}} \sum_{i=1}^{n_{SV}} y_i \alpha_i \left( \vec{x}_i\cdot\vec{x} \right)^n + b~,
\end{equation}
where now $i$ labels the support vectors, their number being
$n_\text{SV}$. The value of the decision function $d(\vec{x})$ (whose
general form is provided in Eq.~(\ref{eq:dfgeneral})) is the
signed distance of the configuration $\vec{x}$ in the image space of
the feature map from the the maximum margin
hyperplane. If we pick an ensemble of independent configurations
$\left\{\vec{x}\right\}_T$ at temperature $T$, the average of the
decision function over these, $d(T) = \langle d(\vec{x})\rangle_T$,
is a thermodynamic observable that depends on $T$. 

In order to find whether a second order phase transition occurs and its
location, two criteria will be required from the model obtained by
training a SVM. First, for an appropriate choice of kernel, the SVM must be able to separate
configurations drawn at any pair of sufficiently separated
temperatures.\footnote{Sufficient separation can be defined in terms of
standard deviations of a chosen thermodynamic observable. Here we will
not need to develop further this intuitive concept.} We say
that the SVM is able to \emph{separate} configurations at two
different temperatures if a maximum
margin separating hypersurface can be found that does not
overfit. Whether this happens can be measured from estimates of the expected risk,
Eq.~(\ref{eq:exp_risk}).  The SVM is able to \emph{separate} configurations at the training
temperatures if the estimated expected risk is small. This ability
will depend on the kernel and on the training
temperatures.\footnote{The value of the regularisation parameter $C$
  will be chosen in each case so that the results do not depend on its
  value. This is not always possible, but in our case results turn
  out to be independent of $C$ provided its value is bigger than $\sim
  10^{2}$. Note that a different choice has been made
  in~\cite{Ponte}.} Second, if trained at two temperatures, the decision
function must be a monotonic function of the intermediate
temperatures. Indeed, if the temperature is progressively changed from
$T_1$ to $T_2$, we expect the configurations collected along the
change to be at the start very similar to those at $T_1$, and to become
progressively more similar to those at $T_2$. We expect this general
behaviour to be reflected in the average value of the decision
function as a function of $T$. This can be restated as the request
that $d(T)$ is a monotonic function of $T$. Heuristically, the reason
for requiring monotonicity is that if we are to define the critical temperature $T_c$ as the
temperature at which $d$ has some set value, for
example when it becomes compatible with $0$, then we must be able to
associate a unique value of $T$ to each value of $d$. Thus $d(T)$ must
be invertible over its domain, and monotonicity follows. Moreover, since
we are analysing different volumes $L$, we will require that the direction
of variation of $d(T)$ does not change qualitatively if $L$ is
changed. These are two independent and necessary criteria that we
require from the model in order to be able to eventually locate the
transition precisely. They will be tested in the next two
subsections. In a third subsection, we will study the meaning of the
decision function that performs best and lastly we will use its
scaling properties to extract the critical temperature $T_c$ and the
critical exponents of the transition. The Machine Learning analyses
reported in this work have been done using the scikit-learn
library~\cite{scikit-learn} and, as a cross-check of results, code
developed in MATLAB.  

Since a priori we do not know the location of the transition, or if
the transition is there at all, we collected configurations at
temperatures $T_i = 0.5 + i 0.5$ with $i=0,\dots,9$ and for the
same values of $L$ as in the standard analysis above. This rough scan
of the temperature range from $T=0.5$ to $T=5.0$ will be refined
when we extract the critical temperature and the critical exponents. Let
us stress that, at this point, our only knowledge on the system comes
from its raw configurations available at different volumes and
temperatures and the fact that its geometry is an $L\times L$ square
with periodic boundary conditions. In particular, any global symmetry
that could drive the transition is assumed to be unknown to us.

\subsection{Monotonicity of the decision function}
For this analysis, we train the SVM at the most distant temperatures
in our range, $T_1=0.5$ and $T_2=5.0$ with homogeneous
polynomial kernels of degree $n=1,\dots,4$, and we compute the value of
the average value of the decision function Eq.~(\ref{eq:decfun_poly})
on sets of $200$ configurations collected at intermediate values
$T_i$. For convenience, we define a translated and rescaled decision function
$\tilde{d}(\vec{x})$ as
\begin{equation}
\tilde{d} (\vec{x})= \tfrac{1}{2}
d(\vec{x})-b \ . 
\end{equation}
The range of $\tilde{d}$ is $[0,1]$. If $T_1$ and $T_2$ are in two
different phases, then $\tilde{d}$ resembles an order parameter.
 
\begin{figure}[tbp]
\centering
  \subfloat{\includegraphics[scale=1.]{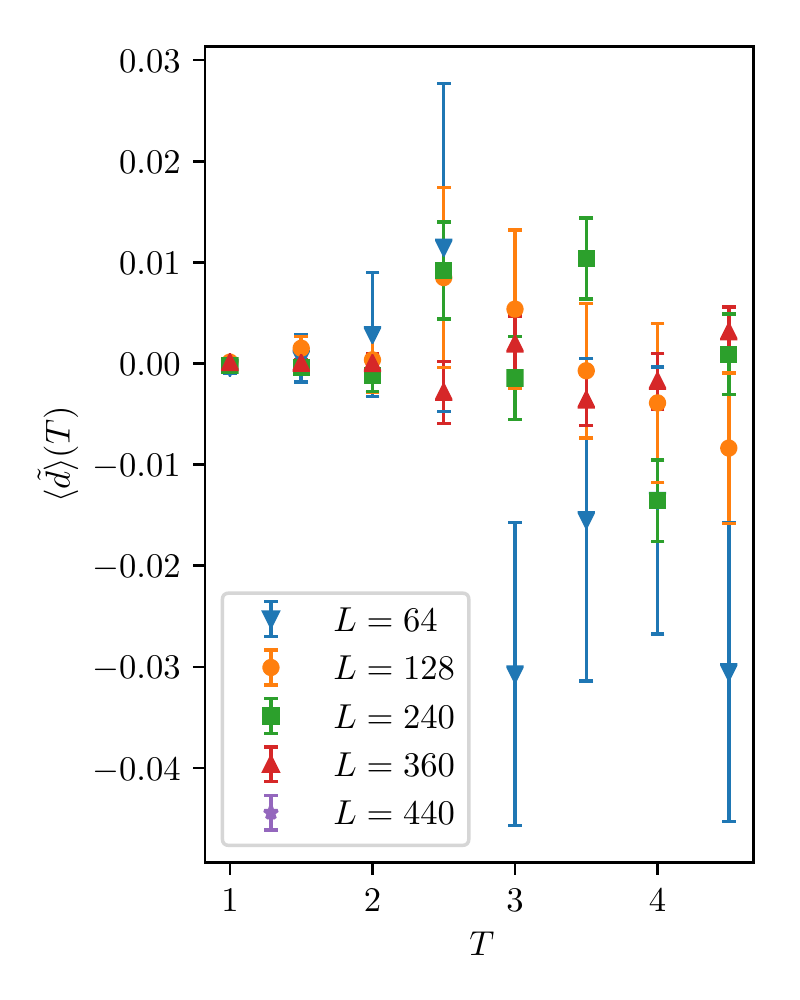}}
  \subfloat{\includegraphics[scale=1.]{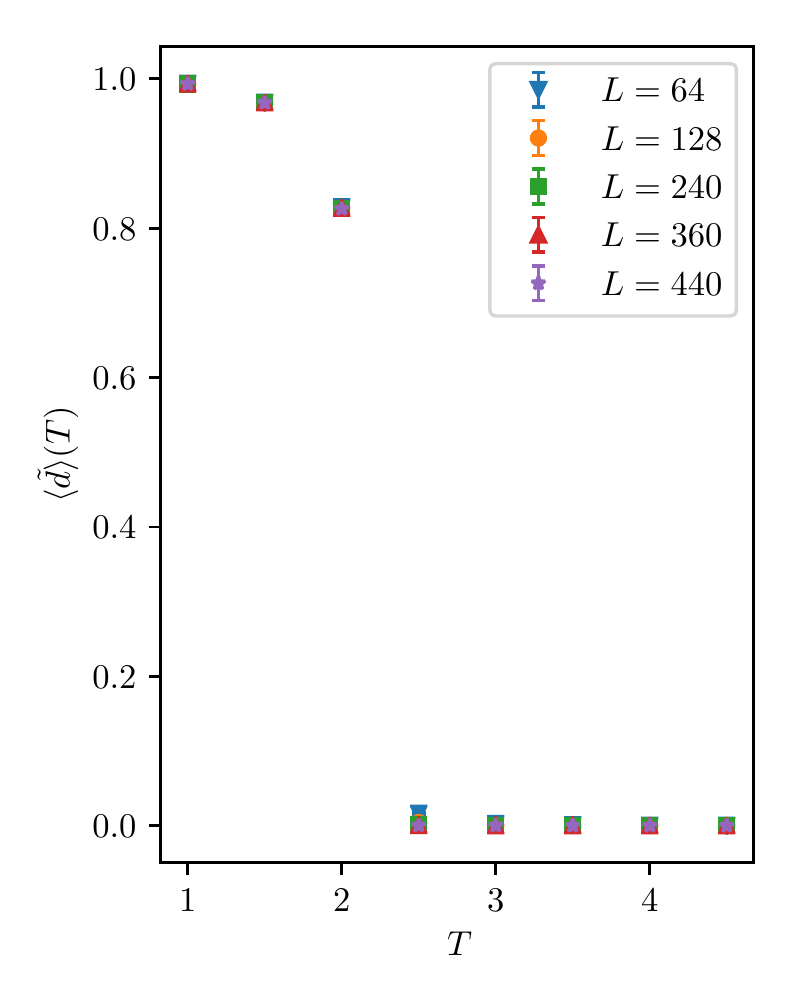}}
  \caption{(Colour online) Behaviour of $\langle\tilde{d}\rangle(T)$ with a kernel of degree $n=1$ (left) and $n=2$ (right).}
  \label{fig:SVM_prelim}
\end{figure}

The values of $\langle\tilde{d}\rangle_T$ are reported
in Fig.~\ref{fig:SVM_prelim}. In the left panel we report the
results for the $n=1$ polynomial kernel, on the right the results for the $n=2$
polynomial kernel. The results do not show any qualitatively appreciable variation as long
as $C>10^{-2}$ and are represented just for $L<440$ in order to avoid 
overcrowding the plots. At larger volumes, their qualitative behaviour does
not change. 

For the $n=1$ polynomial kernel SVM, the classification at intermediate
temperatures is not obviously monotonic (in fact, a classification
signal seems to be completely absent), and, for the same $T$ it also changes
drastically for different $L$'s. Both these features can be
consequences of a rather noisy nature of the hyperplane finding
process in this specific case, where the data for
$\langle\tilde{d}\rangle_T$ are mostly compatible with zero.  Although
this type of kernel fails already at this stage and in such a
spectacular manner, we will not discard it for the time being and
postpone any further comment about this and other odd power
kernels to the next subsection. On the contrary, $n=2$ shows the expected monotonic
behaviour for all the values of $L$.  In this case, $\langle\tilde{d}\rangle_T$  is $\sim 1$ at very low $T$
and goes to $0$ for large $T$. The figure shows how the vanishing
of $\langle \tilde{d} \rangle_T$ is concentrated around
$T = 2.5$. Besides, the results are unchanged if other different
training temperatures are considered, i.e. $T_1=1.0$ and
$T_2=4.0$ (a systematic scan of possible pairing of training
temperatures will be done in the next subsection). Note, moreover, that the value of
$\langle\tilde{d}\rangle(T)$ taken at $T=2.5$ goes to $0$ if $L$ is
increased, while far from this value of $T$, it changes minimally. We
interpret this as evidence that around the value $T = 2.5$, the
behaviour of the system changes in a way that is relevant for the
phenomenon we want to observe. This is a first
sign that the transition might be in a region of $T$ around $T =
2.5$. Henceforth, the neighbourhood of
$T = 2.5$ will be called the~\emph{critical region}. 

The study of kernels of order $n=3$ and $n=4$ does not add new insights,
with the $n= 3$ kernel being similar to the $n = 1$ case and
$n=4$ resembling $n = 2$. A clear pattern starts to emerge that shows a
separation between even-order and odd-order kernels. This will be even
more evident in the next subsection. 

\subsection{Separation ability}
In this subsection, using different degree polynomial kernels, we
evaluate the ability of the SVM to separate data through estimates
of the expected risk, as explained in Sect.~\ref{sect:svg}. Specifically, we will evaluate the empirical risk for
homogeneous polynomial kernels of degrees $n=1,\dots,4$, for every
value of the size $L$ and for every possible pair of training
temperatures $T_1$ and $T_2>T_1$ in the coarse temperature scan discussed
above.  

\begin{figure}[tbhp]
  \subfloat{\includegraphics[scale=1.]{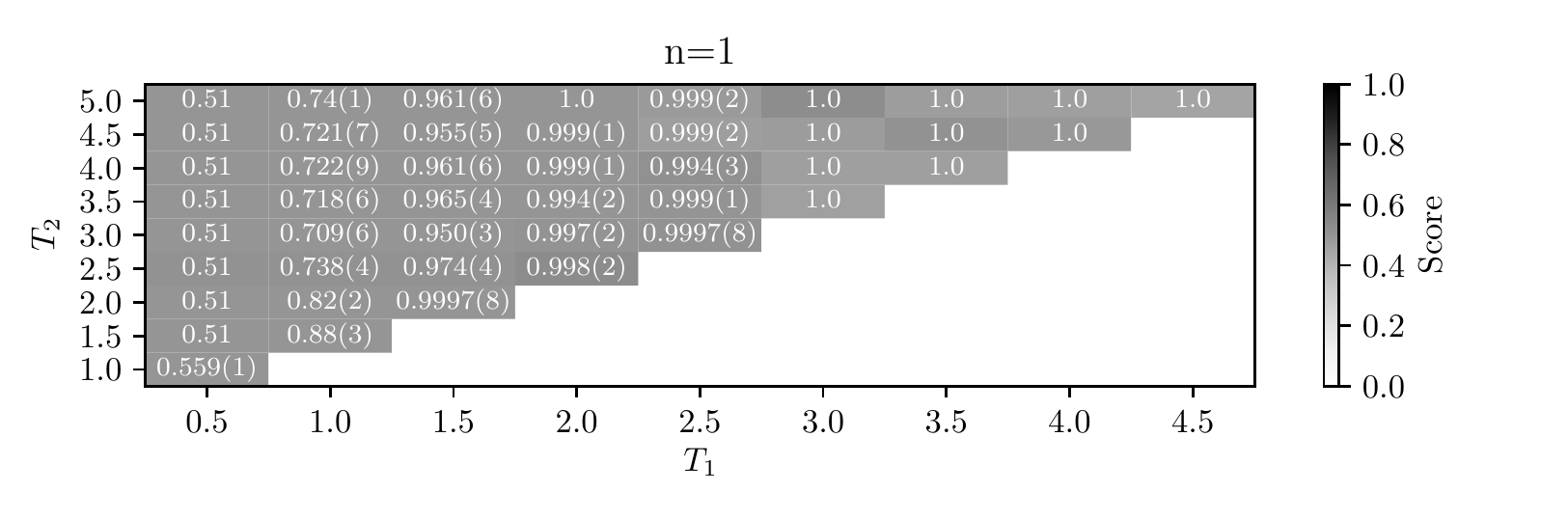}}\\
  \subfloat{\includegraphics[scale=1.]{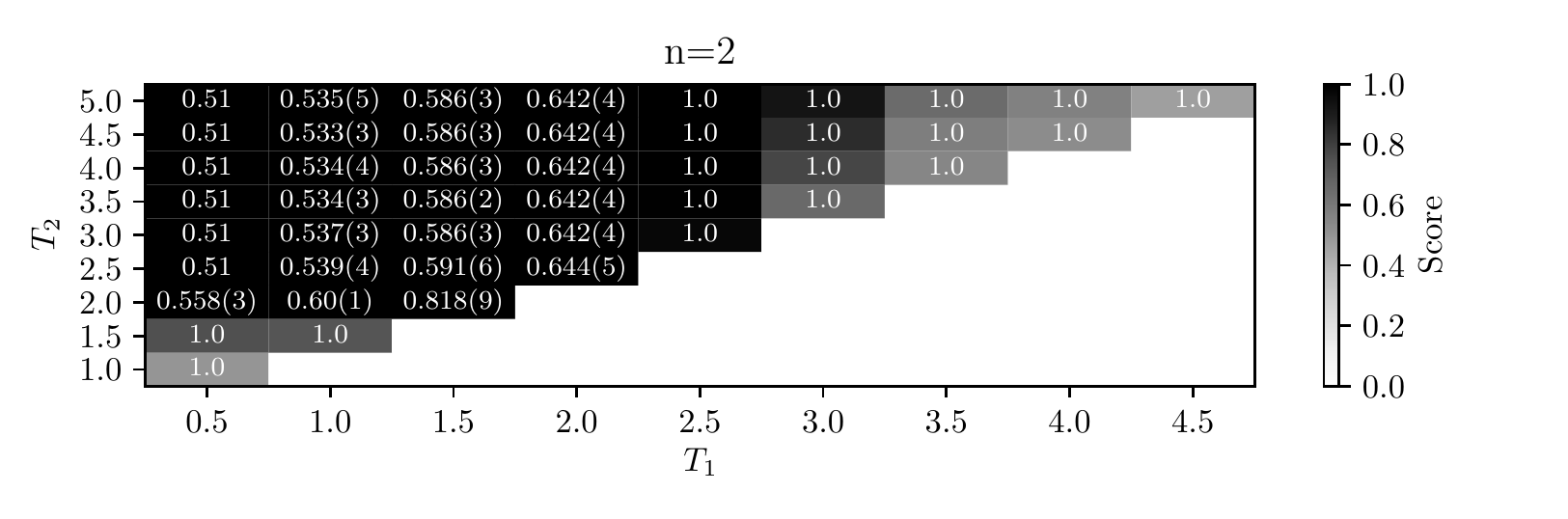}}\\
  \subfloat{\includegraphics[scale=1.]{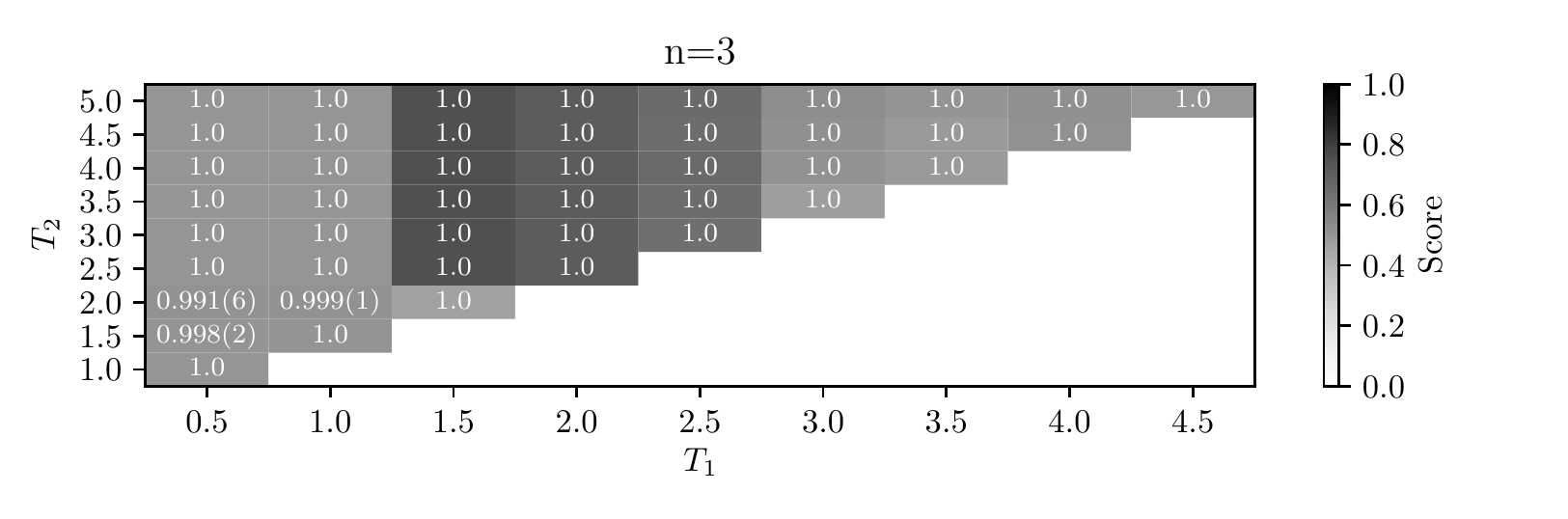}}\\
  \subfloat{\includegraphics[scale=1.]{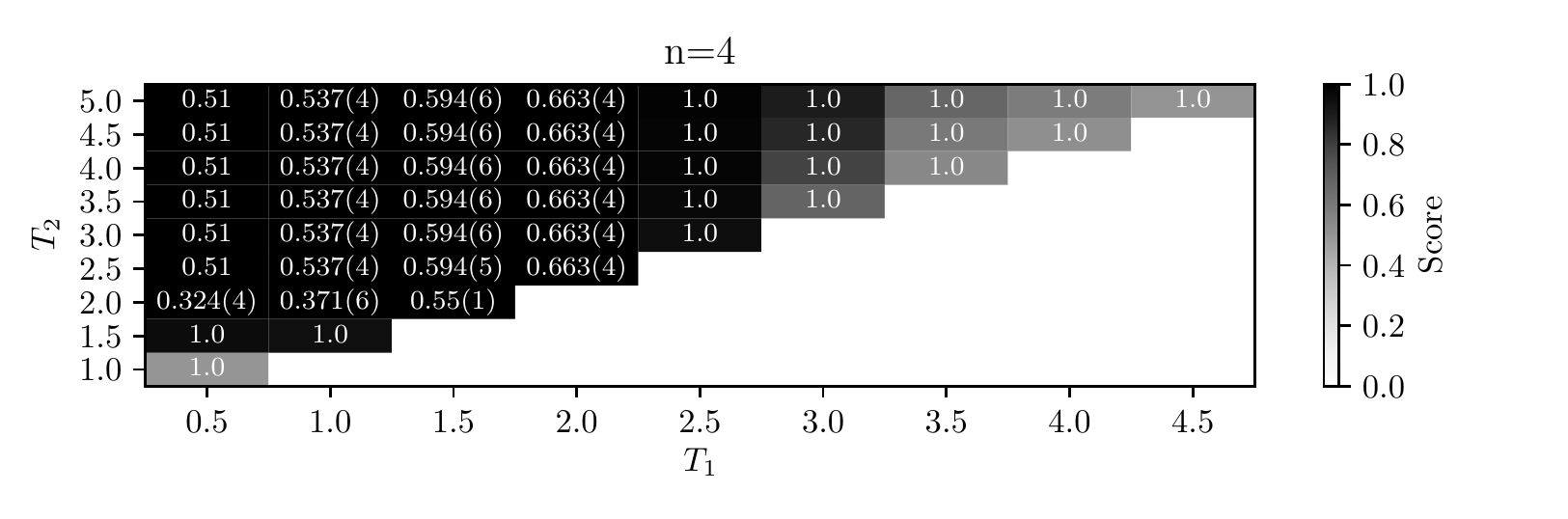}}
  \caption{(Colour online) Heatmaps representing the score of the cross-validation test at $L=128$ and $C=1.0$ with homogeneous polynomial kernels of power $n=1,\dots,4$. The numbers in the rectangles are the estimates of $ \langle n_\text{SV} \rangle/n_\text{TP}$.}
  \label{fig:cross_val_hmap_128}
\end{figure}

\begin{figure}[tbhp]
  \centering
  \subfloat{\includegraphics[scale=1.0]{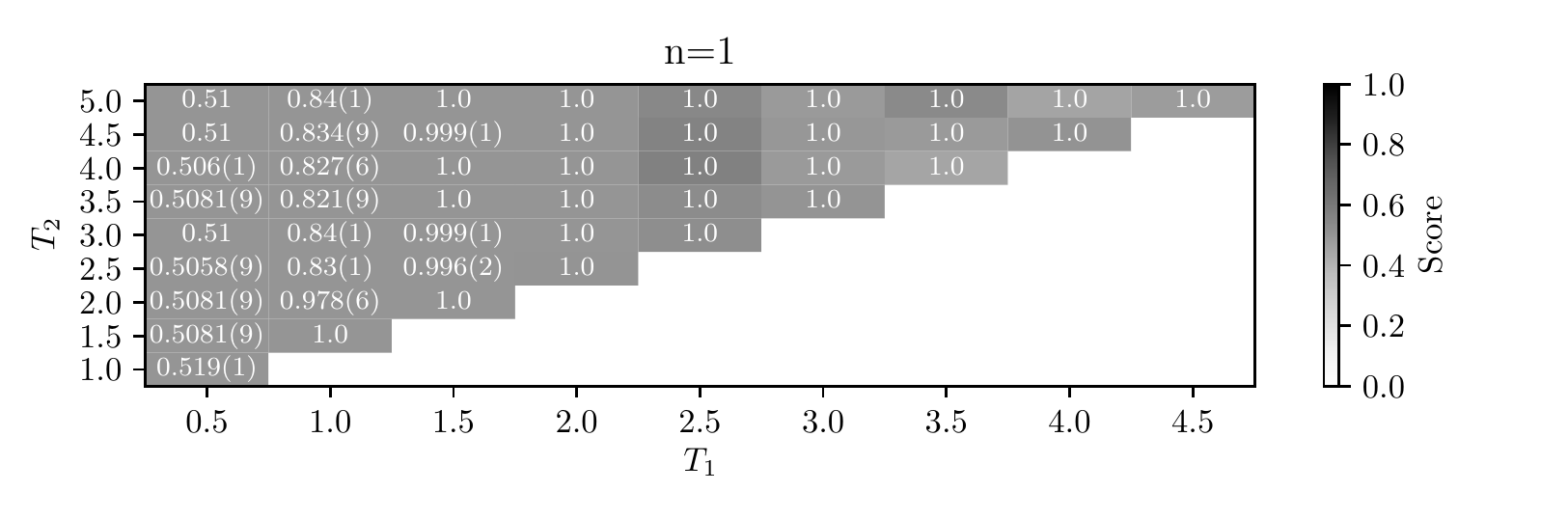}}\\
  \subfloat{\includegraphics[scale=1.0]{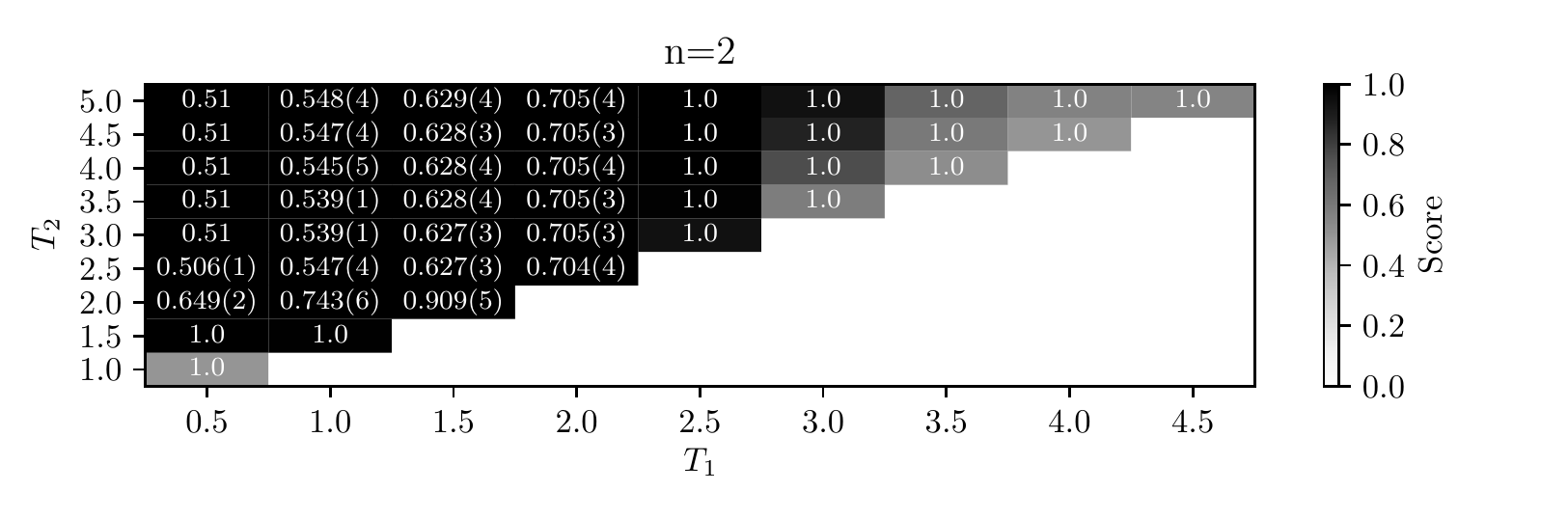}}\\
  \subfloat{\includegraphics[scale=1.0]{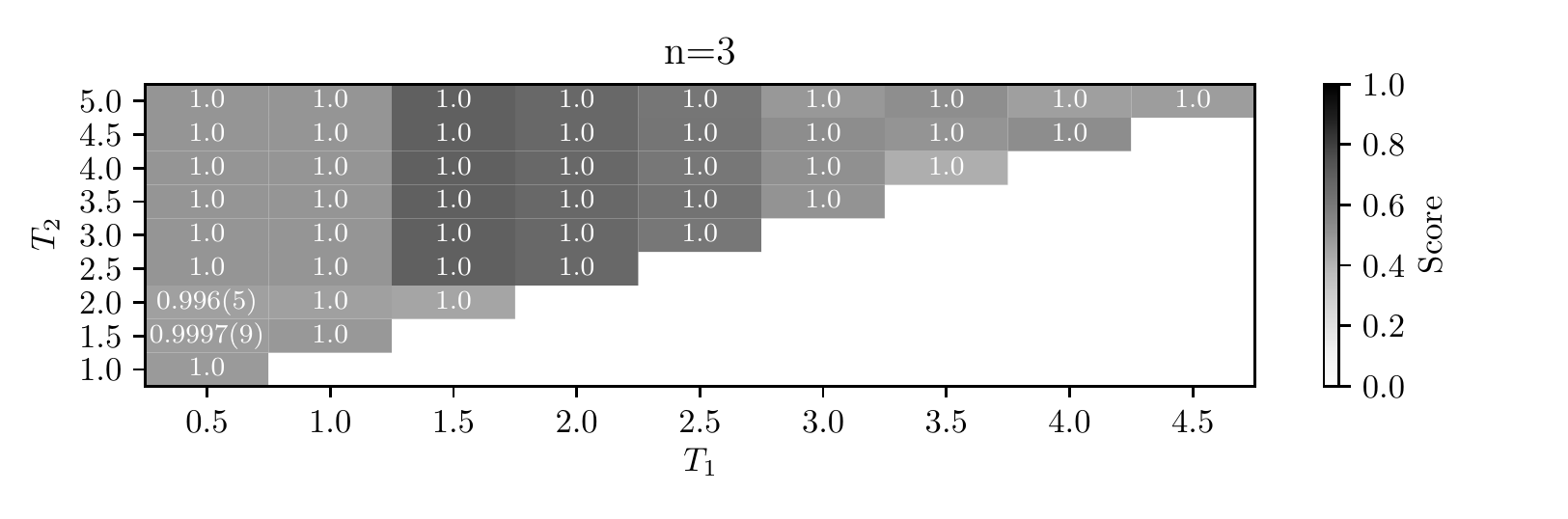}}\\
  \subfloat{\includegraphics[scale=1.0]{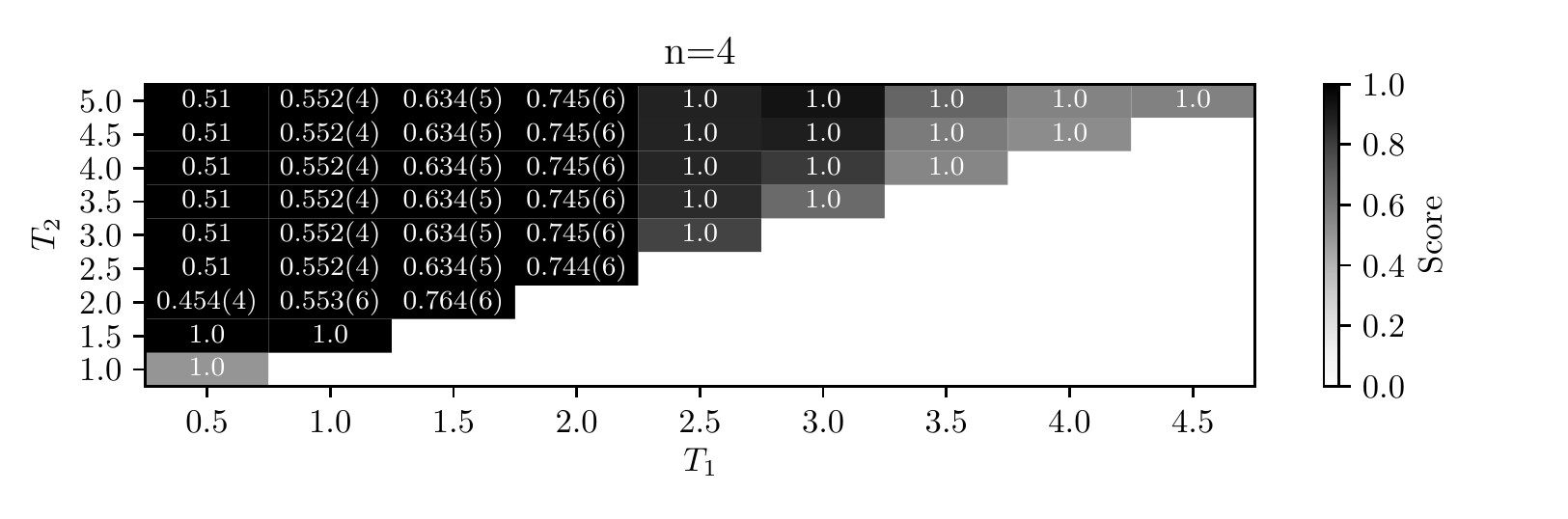}}
  \caption{(Colour online) Heatmaps representing the score of the cross-validation test at $L=240$ and $C=1.0$ with homogeneous polynomial kernels of power $n=1,\dots,4$. The numbers in the rectangles are the estimates of $\langle n_\text{SV} \rangle/n_\text{TP}$.}
  \label{fig:cross_val_hmap_240}
\end{figure}

\begin{figure}[tbhp]
  \centering
  \subfloat{\includegraphics[scale=1.0]{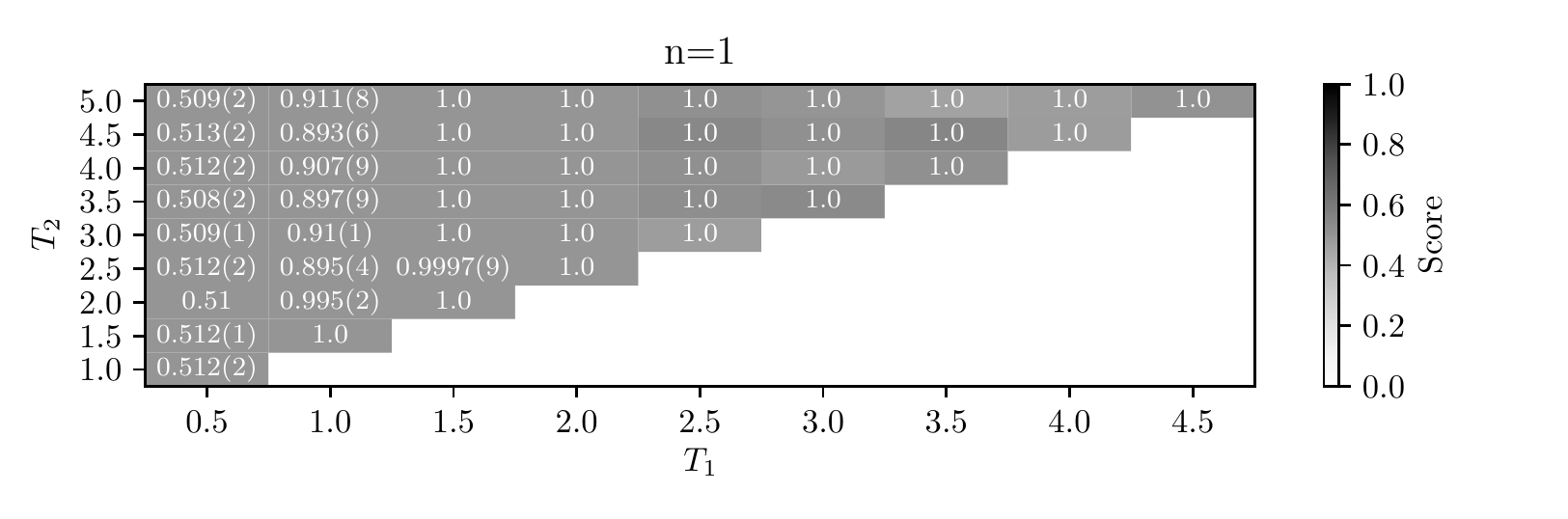}}\\
  \subfloat{\includegraphics[scale=1.0]{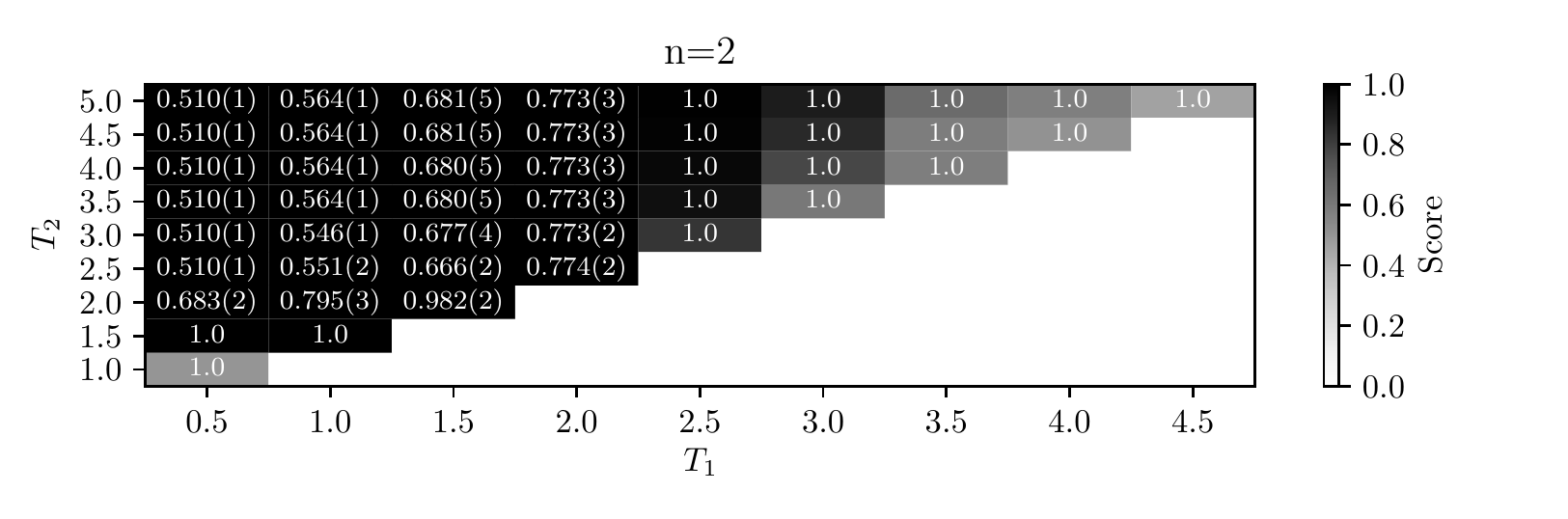}}\\
  \subfloat{\includegraphics[scale=1.0]{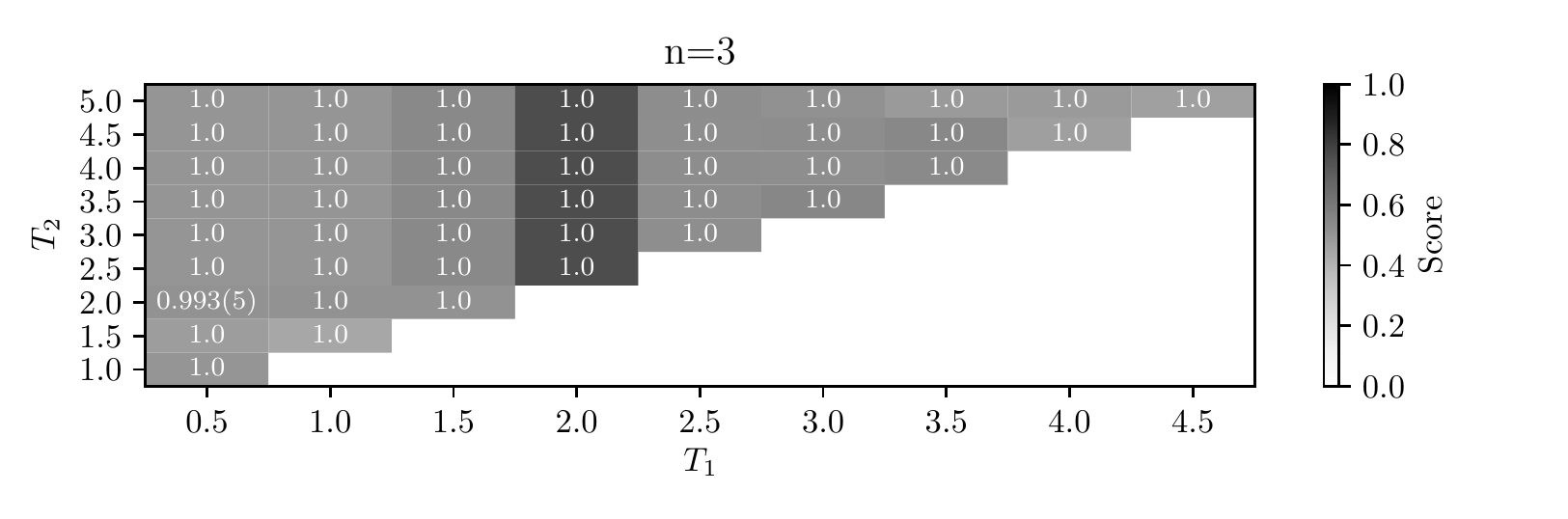}}\\
  \subfloat{\includegraphics[scale=1.0]{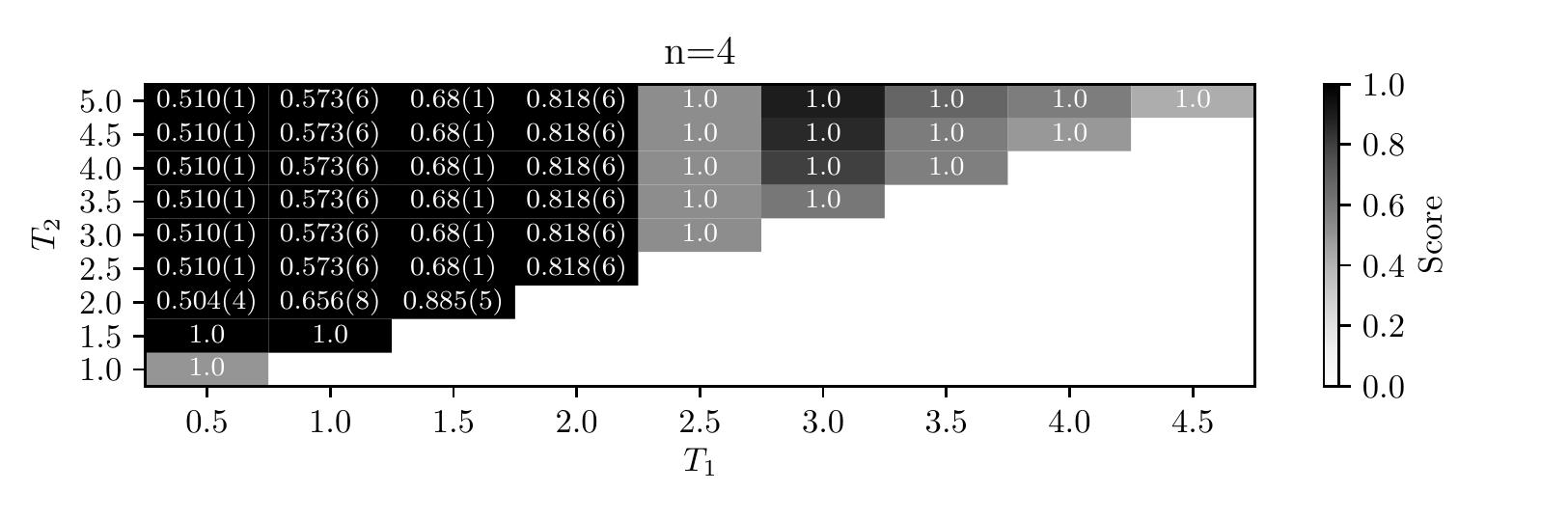}}
  \caption{(Colour online) Heatmaps representing the score of the cross-validation test at $L=360$ and $C=1.0$ with homogeneous polynomial kernels of power $n=1,\dots,4$. The numbers in the rectangles are the estimates of $\langle n_\text{SV} \rangle/n_\text{TP}$.}
  \label{fig:cross_val_hmap_360}
\end{figure}

In each case, we report estimates of the expected risk, also called
the \emph{score}, and of the ratio $\langle n_\text{SV} \rangle/n_\text{TP}$ between number of
support vectors and the total number of points in the training
sample. We estimate the latter with its statistical error using a
jack-knife procedure that consists in computing the running averages
after 10\% of the points in each training set is removed\footnote{
When we perform this estimates we use stratified
sampling, whereby the relative number of configurations in each
class is preserved. In our analysis, configurations have been
ordered according to the Monte Carlo time, i.e., the position at
which they appear in the generated Markov chain. Note that there is no
correlation between the data discarded in each set for the
jack-knife procedure.}; hence, in our case,
$n_\text{TP} = 360$, each complete training set consisting of $200$ points. The results
are represented as heatmaps in Figs.~\ref{fig:cross_val_hmap_128},
\ref{fig:cross_val_hmap_240} and \ref{fig:cross_val_hmap_360} for
three different lattice sizes. In these figures, the lower training
temperature $T_1$ is reported on the horizontal axis, the higher,
$T_2$, on the vertical axis. For each pair, the obtained score is
reported as the color of the corresponding rectangle in grayscale. A
white rectangle maps to the poorest score of $0$, while a black
rectangle to the maximum score of $1$. Note that a score of 0.5 is
equivalent to a mere guess, and hence provides the worst possible
classification ability. In addition, the average of the
ratio $\langle n_\text{SV} \rangle/n_\text{TP}$ is shown as a number in the
rectangle together with the measured error (unless the latter is
exactly zero).  

Let us discuss how to read these heatmaps. In each case, the values
reported on the skew diagonal correspond to close training
temperatures, while, at the opposite end, the values in the upper left
corner correspond to distant pairs of training temperatures. Columns
(resp. rows) correspond to scores obtained for various values of $T_2$
(resp. $T_1$) holding $T_1$ (resp. $T_2$) fixed.  

Already at a first glance we notice that the SVM with a kernel of
even degree ($n=2,4$) yields a better score than with kernels of odd degree
($n=1,3$) everywhere on the heatmaps. Let us analyse the even and the
odd power kernel cases in more detail.  

For the even degree kernels, the results seem to change only slightly
among the various $L$ at $n=2$ and $n=4$. Therefore, we analyse the
case $n=2$, $L=128$ and later comment on the differences with respect to $n=4$
and for larger $L$'s. The score is close to $1$ for almost every
choice of the pair $T_1$ and $T_2$, except when they are both smaller
than $T = 2$ or both greater than $T = 2.5$ \emph{and} very close to each other. Note
that in the critical region, the score remains high even when $T_1$
and $T_2$ are close. Regarding the ratio $\langle n_\text{SV} \rangle/n_\text{TP}$, we can
divide the heatmap in roughly two regions.  
For $0.5 \le T_1 \le 2.0$ and $2.0<T_2\le 5.0$, the ratio has a consistently
lower value than in the rest of the heatmap, the difference being  up
to $\sim30\%$. This is an additional hint at the fact that a
transition may take place for $T\sim2.0$. We remind the reader that
our analysis in the previous section has singled out the value $T =
2.5$, which is the next high up in our coarse scanning; hence, we can
redefine the critical region as $2.0 \le T \le 2.5$. Once again, for
the moment this is just a convenience name, as we have not shown any
evidence of any phase transition yet. 
In the critical region of the heatmap, the ratio $\langle n_\text{SV} \rangle/n_\text{TP}$ reaches its minimum value when
$T_1$ and $T_2$ are the farthest possible, in the upper left hand
corner, and its maximum value when the training temperatures are at their
closest, in the bottom right hand corner. 

In going to $n=4$ the picture is qualitatively the same, as it is for larger values of
$L$. The only difference worth a comment is the behaviour
of the number of support vectors for $T_1=0.5$ and $T_2=2.0$ at $n=4$,
which is systematically lower than the value at $n=2$. As we can observe,
however, the difference reduces at growing $L$ and we interpret it as
an effect of the finite size of the system.  Hence, we can infer that
the estimate of the ratio $\langle n_\text{SV} \rangle/n_\text{TP}$ remains roughly constant
even when the number of  components of the feature map is increased considerably in passing
from a quadratic to a quartic kernel. This means that of the new
components of the feature map, almost none is chosen as support
vector.\footnote{It is worth remarking that, since $\sigma_i^2 = 1$,
  the quartic kernel contains all the terms of the quadratic
  kernel. More in general, a kernel of power $n$ contains all terms of
  kernels of power $m < n$, with $m$ having the same parity as $n$.} This
can be interpreted as the signal of the fact that the $n=2$ kernel
already captures the essential properties of the model. It is then no
surprise that increasing the number of components 
of the feature map does not lead to an improvement of the (already
high) score. In that case, the new components of the feature map are
actually fitting the statistical noise. 

For the odd power kernels $n=1$ and $n=3$, the picture is totally
different, the score being small over all the pairs $T_1$ and
$T_2$. Let us analyse the case $n=1$ for $L=128$ first. The score is
$\sim 0.5$ over all the heatmap. This means that the classification
algorithm classifies incorrectly roughly half of the test samples. In
passing to $n=3$, the score heatmap remains approximately the same, but the
behaviour of the ratio $\langle n_\text{SV} \rangle/n_\text{TP}$ changes, and becomes almost
uniformly $\sim 1.0$. This means that as a consequence of the addition
of components to the feature map, almost the whole points in the
sample become support vectors. Such a behaviour in passing from $n=1$
to $n=3$ is observed at all values of $L$. We argue that in those
cases the minimisation algorithm uses the more numerous components of
the $n=3$ feature map to try to fit to statistical noise, i.e. to
overfit in order to accommodate a separating hypersurface. Hence, in
this case we obtain a poorer classification prediction.  

We have performed a similar analysis (not reported here) also with the
$n = 5$ and $n = 6$ kernels, which confirms the conclusion that even
power kernels are preferred to odd power kernels in terms of their
efficiency at separating classes corresponding to temperatures.  One
can also easily see that, among the even degree kernels, the quadratic
kernel performs better, in the sense of containing already all
information on the separating hypersurface with the minimal number of
features. This set of observations is very powerful at identifying the important symmetry at play. In fact, the common
symmetry of the better performing kernels is $\mathbb{Z}_2$, and among all
the $\mathbb{Z}_2$ symmetric kernels, the $n=2$ kernel is the one for
which this symmetry is maximal, while all the others have higher order symmetries containing
$\mathbb{Z}_2$ as a subgroup (namely, $\mathbb{Z}_4$ for $n = 4$ and $\mathbb{Z}_6$
for $n = 6$). Among the even power kernels, the behaviour of the number of
support vectors singles out the quadratic kernel as the one that best
adapts to the data. 

It is worth stressing again at this point that $\mathbb{Z}_2$
invariance was not an input: indeed the SVM has been only fed with raw
configurations, i.e. vectors with $L^2$ components labeled with the
temperature, whose possible values are $\pm1$. By clearly
singling out the quadratic kernel, the SVM is giving us a strong
hint of a possible global symmetry of the system. Hence, our working
hypothesis that a systematic investigation of a class of kernels can
identify the symmetry of the system seems to be valid. 

\subsection{Meaning of the decision function}
Given the special role played by the $n = 2$ kernel, we restrict our
analysis to the latter from now on. As noted in \cite{Ponte}, in the case of $n=2$, the meaning of the decision function can be easily understood. Its homogeneous part $\tilde{d}$ can be written as
\begin{equation}\label{eq:dec_fun_2_hom}
  \tilde{d}(\vec{x}) = 
  \frac{1}{2}\sum_{i=1}^{n_{SV}} y_i \alpha_i ~\mathcal{K}\left( \vec{x}_i,  \vec{x} \right)  = 
  \frac{1}{2 L^4}\sum_{i=1}^{n_{SV}} y_i \alpha_i ~\left(\sum_{\vec{a}} x_i(\vec{a})\,x(\vec{a})\right)^2~,
\end{equation}
where on the right hand side we switched back to a cartesian labeling
of the elementary variables, $\vec{a}$ indicating the position on the
lattice and the sum running over the whole lattice. After swapping the
sums over the positions\footnote{In the quadratic kernel, there
  are two sums over positions to be performed.} with the sum over the
support vectors, Eq.~(\ref{eq:decfun_poly}) can be rewritten as
\begin{equation}\label{eq:dec_fun_2}
  \tilde{d}(\vec{x}) = \frac{1}{2}\sum_{\vec{a},\vec{b}} C(\vec{a},\,\vec{b}) \, x(\vec{a})\,x(\vec{b})~,
\end{equation}
where
\begin{equation}
  \overline{C}(\vec{a},\,\vec{b}) =
  \frac{1}{L^4}\sum_{i=1}^{n_\text{SV}} y_i
  \alpha_i\,x_i(\vec{a})\,x_i(\vec{b}) \ .
\end{equation}

\begin{figure}[htbp]
  \subfloat[$T_1 = 0.5$, $T_2 = 5.0$.]{\includegraphics[scale=1.0]{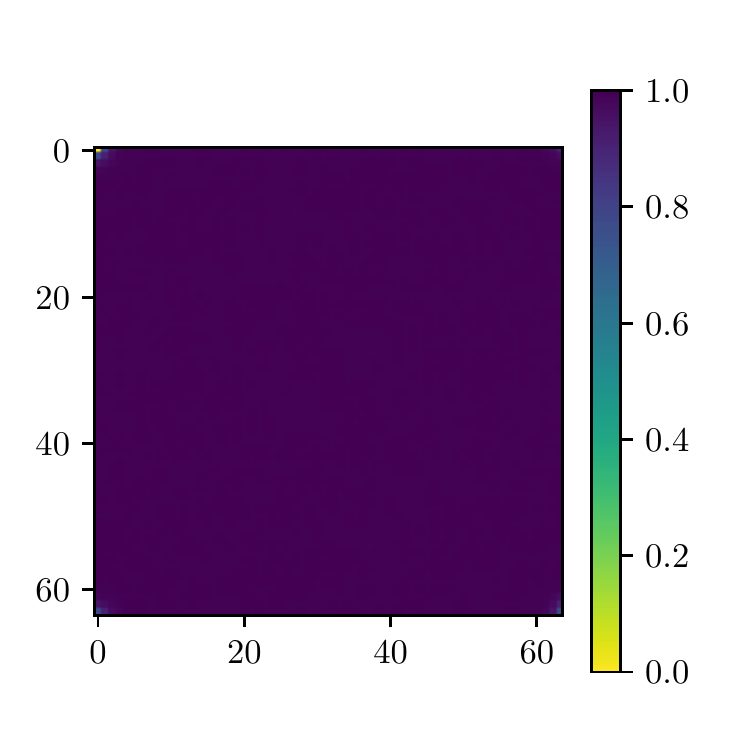}}
  \subfloat[$T_1 = 0.5$, $T_2 = 2.0$.]{\includegraphics[scale=1.0]{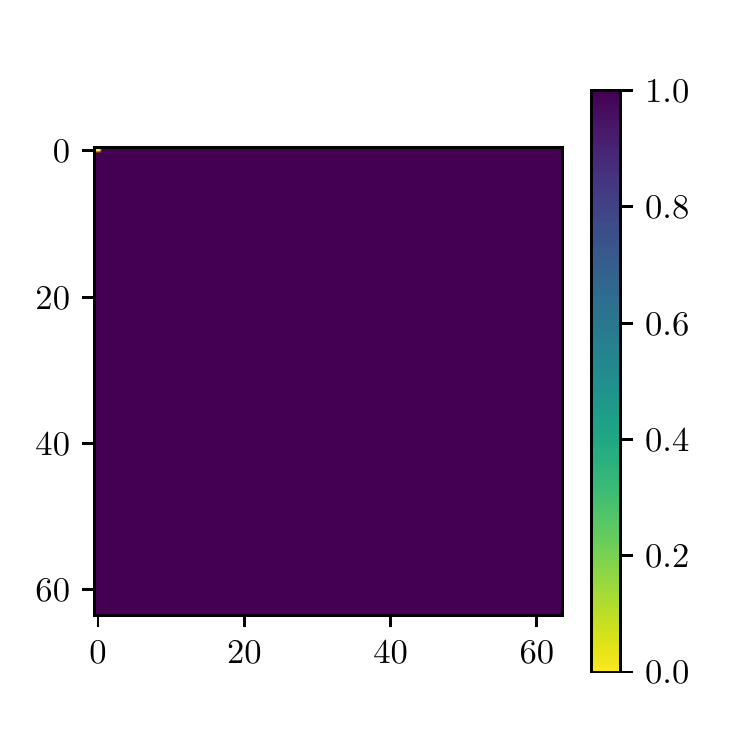}}\\
  \subfloat[$T_1 = 3.0$, $T_2 = 5.0$.]{\includegraphics[scale=1.0]{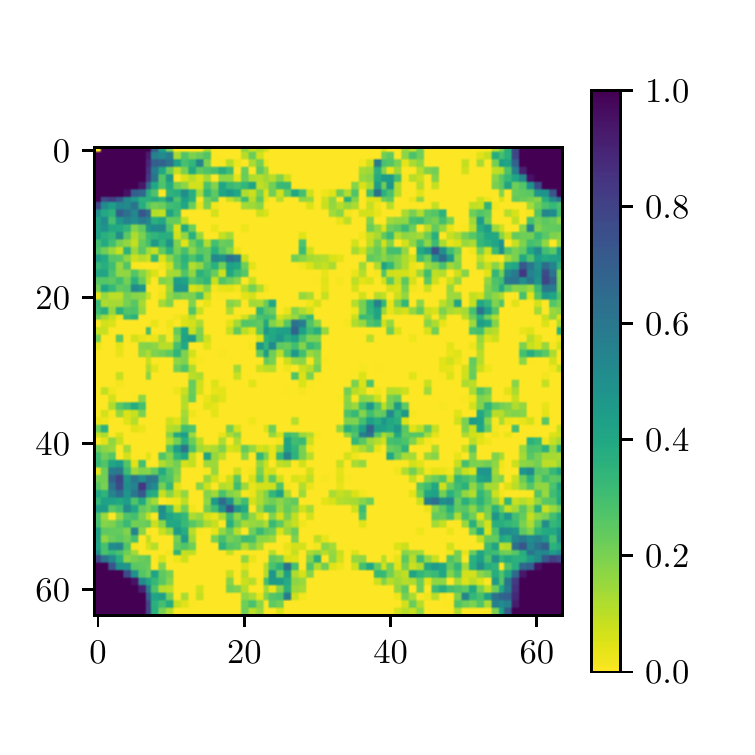}}
  \subfloat[$T_1 = 2.0$, $T_2 = 2.5$.]{\includegraphics[scale=1.0]{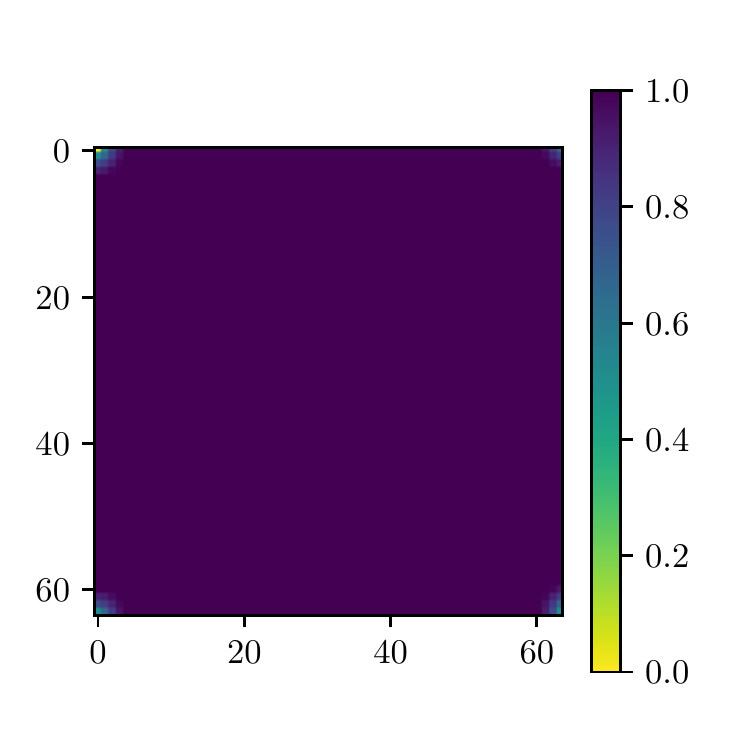}}
  \caption{(Colour online) Heatmaps representing the effective coupling
    $\overline{C}({\vec{x}})$ for $L=64$ at the indicated values of
    the pairs $T_1$, $T_2$. The axes are the Cartesian coordinates of
    the lattice.} 
\label{fig:hmap_decision}
\end{figure}

The quantity $\overline{C}(\vec{a}, \vec{b})$ can be interpreted as an
\emph{effective coupling} between two spins at positions $\vec{a}$ and
$\vec{b}$ (see also~\cite{Ponte}). As it can be verified by direct inspection, $\overline{C}(\vec{a},
\vec{b}) = \overline{C}(\vec{a}-\vec{b})$, owing to the translation symmetry of
the system. By studying the average $\overline{C}(\vec{a})$ obtained by
training a quadratic SVM at a given pair of training temperatures, we
can get a clear insight into the nature of the decision
function. An illustration of the behaviour of this quantity is
provided in a normalised (from $0$
(lightest colour) to $1$ (darkest)) heatmap in Fig.~\ref{fig:hmap_decision} at
$L=64$ for $C=1.0$ at four choices of the training temperatures. These
can be seen as pictures of the effective coupling. Note that periodic
boundary conditions are imposed in both directions. 

Let us comment on these heatmaps. In all cases except one, the effective coupling is roughly \emph{uniform}. Then 
\begin{equation}
  \tilde{d}(\vec{x}) \propto \frac{1}{L^4}\sum_{\vec{a}} x(\vec{a})\,x(\vec{a}) = m^2~,
\end{equation}
where $m$ is the magnetisation density of the system. The decision
function $d$, in this case, is thus linearly related to $m^2$.

If $T_1$ and $T_2$ are respectively smaller and greater than
$T = 2.5$ (bottom right panel) but very close to each other, we see
that the effective coupling vanishes smoothly in a small neighbourhood
of the origin and is uniformly $1$ everywhere else. When, instead,
$T_1, T_2 > 2.5$, the shape of the effective coupling drastically
changes. The decision function becomes now a shorter ranged version of
$m^2$. These conclusions do not change for larger volumes and as long
as $C\gtrsim10^{-2}$. It is clear that the distinction between the
decision functions learned in each case is related to its \emph{range}
on the lattice. The mapping of the decision function into the order
parameter provides us with an easy {\em a posteriori} interpretation
on the conclusion (reached in the previous two subsections) that
learning temperatures must be chosen distant enough and on either side
of $T = 2.5$: with this choice, the SVM has information from both
phases of the model, which enables it to learn the order
parameter.\footnote{The Reader would have noticed that in
  Sect.~\ref{sect:ising} we used as an order parameter $m$, while here we are
  claiming that the order parameter is $m^2$. Indeed, the issue is
  subtle: strictly speaking, the correct order parameter is $m$, since
a request for an order parameter is that it has to transform
non-trivially under the symmetry of the system. However, on a finite
lattice, $m$ is always zero. Hence, the learning process will identify
$m^2$ as a classification function, associating the two phases
respectively with the region of temperatures in which this quantity is
of order one and the region in which it is much smaller than one.}

\subsection{Extracting $T_c$ and the critical exponents}
In this subsection, we shall show that the decision function can be
used to precisely locate the phase transition point and evaluate the
critical exponents. The heuristic argument that guides our approach
is the following. Far from the phase transition, the SVM should
easily succeed at classifying
phases, as configurations will fall far from the separating
hypersurface. Near the phase transition, we would expect a less clear
classification, with configurations falling in the separating
margin. The critical temperature can be identified with the
temperature at which $\langle \tilde{d} \rangle$ has the maximal change
(see e.g. Fig.~\ref{fig:SVM_prelim}). Hence, by studying the fluctuations
of the decision function obtained following an appropriate training
procedure (which is provided by a measure of the classification
error), one can identify  the critical temperature as the value at
which this quantity reaches its maximum. Owing to finite size scaling,
the shift of this maximal value as a function of $L$ is expected to
scale with the critical exponent $\nu$ of the transition. 

For the Ising model, the identification of $m^2$ as the decision function learned for distant
temperatures on either side of $T = 2.5$ provides a more rigorous
justification of our heuristic expectations. Indeed, we know that the
fluctuations of $m^2$ reach their peak at the critical
temperature. Since $\tilde{d}$ is proportional to $m^2$, its
fluctuations should show the same behaviour. Therefore, by
performing a finer scan of temperatures in the critical region as
identified in the procedure for calibrating the choice of the kernel,
we should be able to find a peak in $\sigma_d$, the susceptibility of
$d$, hence uncovering the phase transition. This will allow us to
obtain $T_c(L)$ and the  
critical exponents.  

Before showing our results, we remark that our discussion of the
methodology allows us to identify also important potential sources of 
systematic errors. For instance, when instead of $m^2$ the learned
decision function is a shorter range version thereof, as it happens
e.g. when both training temperatures are in the symmetric
phase, the fluctuations of $\tilde{d}$ will not be related to criticality, and the
outcome of our analysis will be completely dominated by the
systematics. We have shown that our procedure for choosing the
training temperatures reasonably protects us from this extreme
scenario. Other distortions to the learned decision function from the
target one will arise if $T_1$ and $T_2$ are chosen too close to the
critical region, as shown in the bottom right panel of
Fig.~\ref{fig:hmap_decision}. Although cross-validation and 
analysis of number of support vectors as the training temperatures
vary provide reassuring evidence that we can avoid also this
case, our determination of $T_c$ and of the critical exponents must
take it into account as a logical possibility. Hence, while for
brevity we shall show only results for one set of training
temperatures (namely $T_1 = 0.5$ and $T_2 = 5.0$), in our study we ensured we are free from
systematic errors related to the choice of the training points by 
testing our numerical values for robustness against changes of $T_1$
and $T_2$ in the pre-determined acceptable region. 

We now move on to the determination of the critical temperature $T_c$ and
of the critical exponents. In order to
perform an easier comparison with the results obtained with standard
methods for the magnetic susceptibility $\chi$, we consider the
quantity 
\begin{equation}
  V \sigma_d = V \sqrt{ \langle d^2\rangle-\langle d \rangle^2}  \ .
\end{equation}
From Eq.~(\ref{eq:dec_fun_2_hom}), we find
\begin{equation}
 \label{eq:scaling_sigmad}
 V \sigma_d \propto V \sqrt{ \langle m^4 \rangle - \langle m^2
   \rangle^2} = V \langle m^2 \rangle \sqrt{\frac{\langle m^4
     \rangle}{\langle m^2\rangle^2}-1} \ . 
\end{equation}
Note that $\sigma_d$ and $\sigma_{\tilde{d}}$ are linearly related,
owing to the definition of $\tilde{d}$.  

Let us examine the scaling behaviour of $V \sigma_d$. The choice of $\Gamma$ in
Eq.~(\ref{eq:choice_gamma}) ensures that the proportionality constant
between $\tilde{d}$ and $m^2$ is independent of the volume. Hence, a straightforward dimensional analysis
of Eq.~(\ref{eq:scaling_sigmad}) shows that the scaling behaviour of $V \sigma_d$
near criticality is the same as that of the magnetic susceptibility $\chi$, 
\begin{equation}\label{eq:scalingchid}
   V \sigma_{d}(T_c(L)) \propto L^\frac{\gamma}{\nu}~ \ , 
\end{equation}
with 
\begin{equation}\label{eq:scalingmaxchid}
T_c - T_c(L) \propto L^{1/\nu} \ . 
\end{equation}

To obtain both the pseudocritical temperature $T_c(L)$ and the
critical exponents $\gamma$ and $\nu$, it will then be sufficient to
find the coordinates of the maximum of $V \sigma_d(L)$ in the $(T, V
\sigma_d)$ plane (which we refer to respectively as $T_c(L)$ and $V
\sigma_\text{d,max}(T_c)$) for each $L$ and
fit~Eq.~(\ref{eq:scalingchid}) and Eq.~(\ref{eq:scalingmaxchid}) to
their behaviour. For each value of $L$, $V \sigma_\text{d,max}(T_c)$ and $T_c(L)$ are first roughly
estimated and then their estimate is improved with a finer scan. To
better compare the final results with those obtained 
with the multi-histogram method, we used the same temperatures used for
this latter analysis, see Tab.~\ref{tab:betac}. 

\begin{figure}[tbp]
  \centering
    \subfloat[\label{fig:TcvsL-ML} (Colour online) Temperature of the
    peak value of $V\sigma_d$ at each simulated size $L$.]{\includegraphics{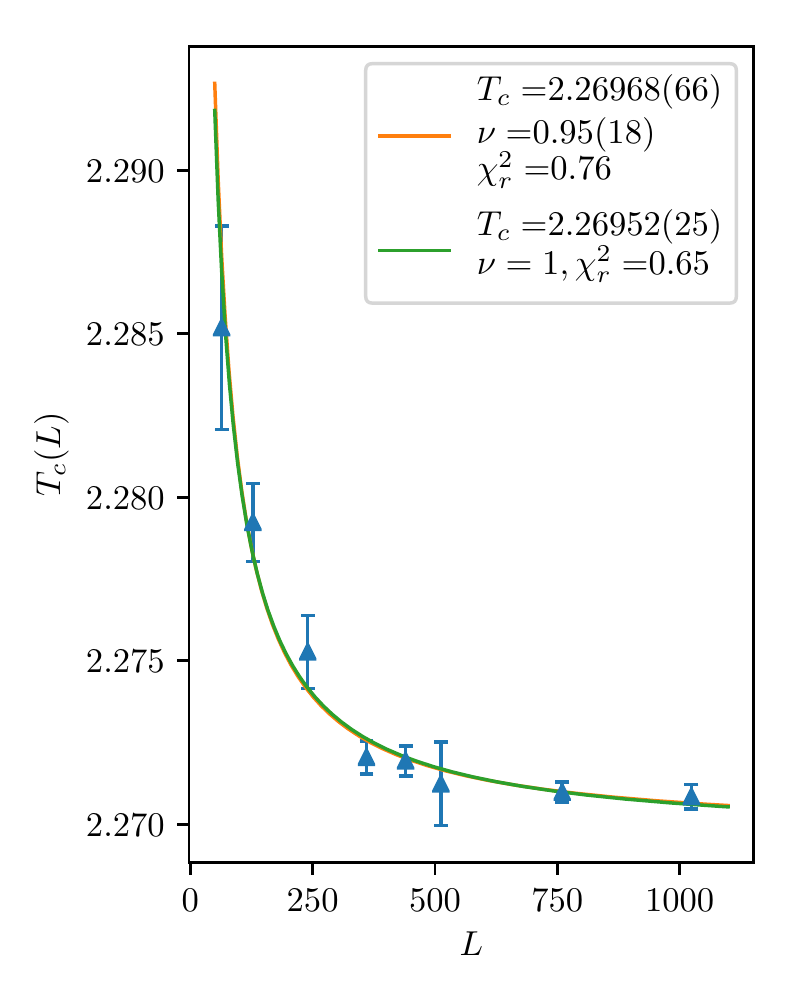}}
  \subfloat[\label{fig:chivsL-ML} (Colour online) Peak values of $V
  \sigma_d$ computed at each simulated size $L$. ]{\includegraphics{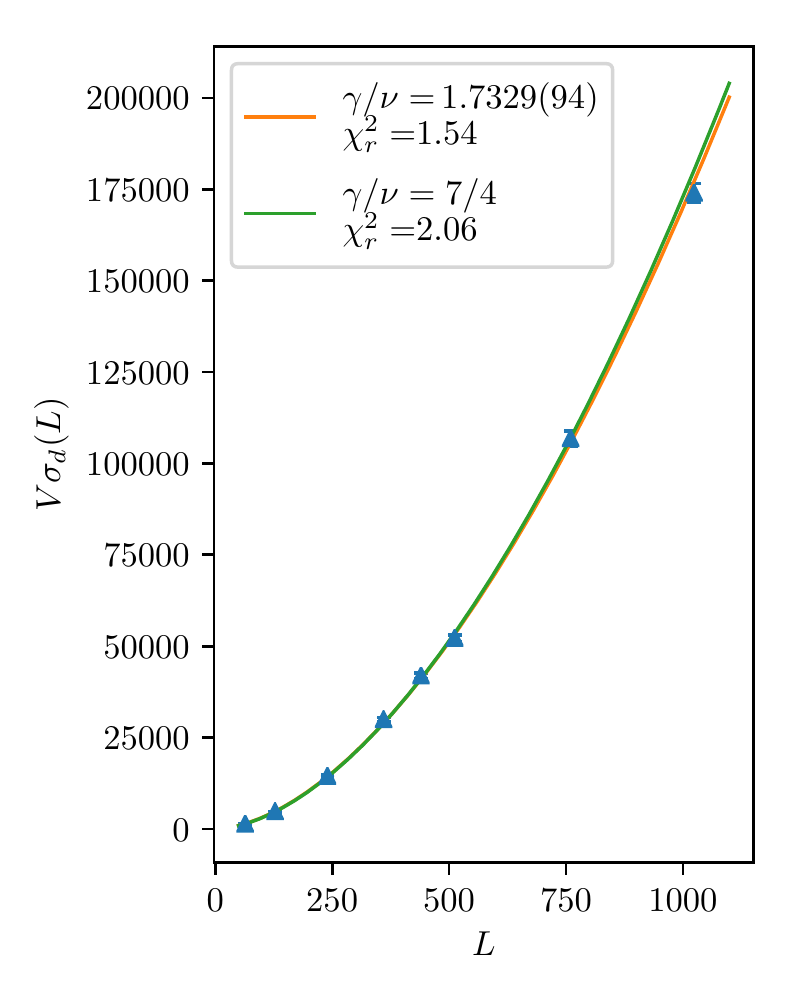}}
\caption{Finite size scaling of critical quantities extracted from the
SVM decision function error.}
\end{figure}

The results of this procedure are reported in Tab.~\ref{tab:betac-ML}
and represented in Figs~\ref{fig:TcvsL-ML}~and~\ref{fig:chivsL-ML}. The scaling behaviour is fitted to the data using
$T_c(L)$, $\nu$ and $\gamma/\nu$ as fitting parameters. The results of
the fit are reported in Tab.~\ref{tab:fits-ML} and plotted also in
Figs.~\ref{fig:TcvsL-ML}~and~\ref{fig:chivsL-ML}. As an additional
estimate of the critical temperature, also the fits with the critical
exponents fixed to their analytical values are performed. The results
are visible in the same figures and tables. The determined values of
$T_c$, $\nu$ and $\gamma$ have
good accuracy, which allows us to make meaningful comparisons with both 
the values obtained analytically for the 2D Ising model ($\gamma=7/4 =
1.75$ and $\nu=1$) and with the estimates obtained with the
multi-histogram method. As in the conventional analysis, all the
determined quantities are compatible with the analytical known values
within at most two standard deviations.  
The errors on the fitted parameters obtained with the traditional
approach are smaller by about a factor two to four than in the SVM analysis, possibly
owing to the fact that the former method combines samples at different
values of $T$ through multi-histogram reweighting, which can not be
used in our Machine Learning analysis (since, for instance, in order
to use it, we would need to know the Hamiltonian of the system, which
is not part of our hypotheses). Likewise, the availability
of more data for the fitting procedure generated through reweighting explains the smaller
$\chi^2_r$ in the case of the conventional analysis, since in this
latter case one has better resolution around the maximum. 

Hence, to conclude, when applied to the same set of input data, our
analysis shows that a finite size scaling study of the peak of the decision
function susceptibility provides results that are quantitatively comparable for precision and
accuracy to those obtained with a traditional finite size scaling
analysis of the order parameter susceptibility using reweighting techniques. 
We note that our analysis for the extraction of $T_c$ and $\nu$ has only made
use of an implicit connection between the decision function and the
order parameter, while for the extraction of $\gamma/\nu$ the exact
relationship has been needed. 

\begin{table}
\begin{center}
  \begin{tabular}{lll}
\hline
$L$ & $T_c (L)$ & $V\sigma_d$ \\
\hline
$ 64  $ & $ 2.2852(31) $ & $  1.426(31) \cdot 10^3 $ \\
$ 128 $ & $ 2.2792(12) $ & $  4.782(85) \cdot 10^3 $ \\
$ 240 $ & $ 2.2753(11) $ & $  1.448(24) \cdot 10^4 $ \\
$ 360 $ & $ 2.27204(51) $ & $  2.995(55) \cdot 10^4 $ \\
$ 440 $ & $ 2.27194(46) $ & $  4.193(82) \cdot 10^4 $ \\
$ 512 $ & $ 2.2712(13) $ & $  5.221(87) \cdot 10^4 $ \\
$ 760 $ & $ 2.27098(31) $ & $  1.068(21) \cdot 10^5 $ \\
$ 1024 $ & $ 2.27085(38) $ & $  1.740(26) \cdot 10^5 $ \\
\hline
\end{tabular}
\end{center}
\caption{Position ($T_c(L)$) and volume-multiplied value
  ($V\sigma_d$) of the maximum of the decision function error at each
  investigated lattice size $L$.} 
\label{tab:betac-ML}
\end{table}

\begin{table}
\begin{center}
	\begin{tabular}{ccc|cc}
		\hline
		$T_c$ & $\nu$ & $\chi^2_r$ & $\gamma/\nu$ & $\chi^2_r$ \\
		\hline
		$2.26968(66)$ & $0.95(18)$ & $0.79$ & $1.733(10)$ & $1.54$ \\
		$2.26954(25)$ & $1$ (exact) & $0.65$ & $7/4$ (exact) & $2.06$ \\
		\hline
	\end{tabular}
\end{center}
	\caption{Results of the fit of the predicted scaling behaviour
          Eq.~(\ref{eq:scalingchid}) and Eq.~(\ref{eq:scalingmaxchid}) to
          the data in Tab.~\ref{tab:betac-ML}. The best fit
          curves are represented in Figs.~\ref{fig:TcvsL-ML}~and~\ref{fig:chivsL-ML}.}
	\label{tab:fits-ML}
\end{table}

\section{Conclusions and outlook}
\label{sect:conclusions}
In this work, we have provided the first (to the best of our
knowledge) precision test of Machine Learning techniques applied to
the study of phase transitions in statistical systems. In particular, we
have studied the Ising model, benchmarking our findings with more
consolidated numerical approaches that assume the knowledge of the
Hamiltonian and of the order parameter. As a Machine Learning tool, we
have used the Support Vector Machine, which implements a supervised
learning technique. Our starting point are sets of
configurations (with phase not specified, or unlabelled, using a
Machine Learning language). The first task has been to understand
whether a phase transition takes place. In order to perform this task,
we needed to optimise the Machine Learning process by choosing a kernel
to map input data in a space in which they are linearly
separable. Looking at the performance of the separation process by
choosing two arbitrary temperatures and giving them two different
labels, we have been able to optimise the kernel and to deduce where a
phase transition takes place. Our procedure of iterating over ordered
pairs of training temperatures can be seen as a way to perform unsupervised learning
using a supervised learning tool. Note that our approach is different
from that of~\cite{Ponte}, where the phase was assumed to be known at
various temperatures. In our case, the phase of the system is an
output. 

The optimised decision function, which is the learned classification
criterium  and is obtained through a systematic
study of kernel performances using the optimal training temperatures,
turns out to be simply related to the order parameter, with the
kernel selection process and the optimal kernel pinning 
down the symmetry driving the phase transition. With the knowledge of
the decision function, we have performed a finite size scaling
analysis of its susceptibility (related to the classification error,
expected to be maximal at the phase transition), obtaining results for
the critical temperature and the critical exponents that are comparable in
precision to those extracted with the best numerical tool currently available for studying
phase transitions in systems with known Hamiltonian, namely finite
size scaling of the order parameter susceptibility. Our extraction of
the critical temperature and of the critical exponent $\nu$ describing
the divergence of the correlation length has relied on the sole
knowledge of the decision function and on the assumption that the latter is
related to the order parameter, but not on the precise relationship
between the two. This explicit relationship has been exploited to determine the
combination $\gamma/\nu$. 

Our results pave the way to precise quantitative studies of phase
transitions using Machine Learning techniques, which are particularly
useful in cases in which an order parameter is either not known or not
existing, such as for topological phases of matters. There are several
related directions in which this work can be extended. First, one can
check whether the method of kernel selection we have proposed works in the
Potts model, where the transition is driven by a $\mathbb{Z}_N$
symmetry, with $N \ge 3$. For $N = 3$, we still expect a second order phase
transition, but now the best performing kernel should be an
homogeneous polynomial of order 3, with homogeneous polynomial of
order 3n ($ n > 1$) still giving similar performance, while
polynomials of order $3n + 1$  and $3n + 2$ ($n \ge 0$) should have significantly worse performance. In
addition, for $N > 4$, the system has a first order phase
transition. Hence, in these cases it is not clear {\em a priori} if we
can use the same methodology we have successfully devised for
a second order phase transitions. Explorations in these directions
are currently in progress. Another relevant question is how the proposed procedure can be
generalised to systems with continuous symmetries, for which we would
need to optimise the kernel in a wider space. Finally, it will be
interesting to test our methodology on systems where a {\em bona fide} order
parameter is absent or not known, like in models of topological superconductivity or
in QCD with finite quark mass.

\section*{Acknowledgements}
We thank Andreas Athenodorou, Claudio Bonati, Michele Caselle, Massimo D'Elia,  Philippe de
Forcrand, Francesco Negro and Enrico Rinaldi for insightful
discussions. CG is partly supported by the EPSRC UKRI Innovation
Fellowship EP/S001387/1. The work of BL is supported in part by the Royal
Society, the Wolfson Foundation and by the STFC Consolidated Grant
ST/P00055X/1. DV acknowledges support from the INFN
HPC\_HTC project. Numerical simulations have been performed on
facilities provided by the Supercomputing Wales project, which is
part-funded by the European Regional Development Fund (ERDF) via Welsh
Government.    

\appendix
\section{Methodology}
\label{sect:appendix}
In this appendix, we discuss more technical aspects of the data
analysis performed in Sects.~\ref{sect:ising}~and~\ref{sect:results}. 

\subsection{Multi-histogram reweighting}
\label{sect:app:reweighting}
In addition to providing an efficient way for computing thermodynamic
observables, Monte Carlo simulations give direct information on the
density of states $\rho(E)$ of a system, which can be extracted
from an energy histogram of the generated configurations at a
particular value of $\beta$. If $n(E)$ is
the number of recorded events at energy $E$ and $N$ is the total
number of generated events, the measured probability for the
occurrence of energy $E$ is 
\begin{equation}
p(E) = n(E)/N \ .
\end{equation}
Since Monte Carlo are first principle methods, in the large $N$ limit
this has to be equal to the Boltzmann probability. Hence,
\begin{equation}
\rho(E) = \frac{Z}{N} n(E) e^{\beta E} = \frac{n(E)}{N} e^{\beta E -
  f} \ , 
\end{equation}
where $f = \beta F$ and $F$ is the free energy of the system. 

In principle, determining $\rho(E)$ from a single simulation performed
at a particular $\beta$ allows us to compute $Z$ (and then, to extract
the thermodynamic properties of the system) at any other value of the
temperature, since
\begin{equation}
Z(\beta^{\prime}) = \sum_E \rho(E) e^{- \beta^{\prime} E} \ .
\end{equation}
The approach of reconstructing thermodynamic observables at different
$\beta$ from the density of states measured with histograms obtained
in a single simulation is called
{\em single-histogram reweighting}~\cite{Ferrenberg0}. 

In practice, however, given that $E$ has Gaussian fluctuations around its average,
in a simulation involving a finite set of configurations $\rho(E)$
can be extracted only in a limited range around the average energy,
since the entries in the histogram will be unavoidably zero far enough
from the central value of the Gaussian. On the other
hand, this very same fact tells us that only a limited number of
states with energy sufficiently close to the ensemble average
Hamiltonian contribute in practice to the thermodynamics of a system
at a given value of $\beta$. In order to cover the relevant range of
energies needed at a particular temperature, one could
do simulations at different values of $\beta = \beta_1, \dots,
\beta_i, \dots, \beta_j$  for which the target density of states provides a
non-negligible contribution to thermodynamic averaging of
observables. For each of the simulated $\beta_i$ and fixed value of
the energy $E_k$, we have 
\begin{equation}
\rho_i(E_k) = \frac{n_i(E_k)}{N_i} e^{\beta_i E_k -  f_i} \ , 
\end{equation}
with $\rho_i(E_k)$ being the density of states at energy $E_k$ measured
in the run at $\beta_i$. Since all values of $\rho_i(E_k)$ are an estimator for
$\rho(E_k)$, we can build the improved estimator
\begin{equation}
\rho(E_k) = \sum_i r(i) \rho_i(E_k) \ , 
\end{equation}
with the weights $r(i)$ satisfying $\sum_i r(i) = 1$. The $r(i)$ can
be determined by minimising the square of the error in $\rho(E_k)$,
which gives
\begin{equation}
\label{eq:multirho}
\rho(E_k) = \frac{\sum_{i=1}^{j} g_i^{-1} n_i(E_k)} {\sum_{i=1}^j N_i
  g_i^{-1} e^{\beta_i E_k - f_i}} \ , 
\end{equation}
with the $f_i$ defined self-consistently using the relationship
\begin{equation}
\label{eq:multiz}
e^{- \beta_i f_i} = \sum_k \rho(E_k) e^{- \beta_i E_k} \ , 
\end{equation}
$n_i(E_k)$ the number of entries at energy $E_k$ recorded in the run
performed at $\beta_i$ and $N_i$ the total number of configurations
generated in the same run. In order to keep into account
the autocorrelation time of each simulation, we have introduced the
autocorrelation factor $g_i = 1 + 2 \tau_i$, where $\tau_i$ can be
calculated e.g. with the Madras-Sokal algorithm~\cite{Madras1988}. The set of
$2j$ simultaneous equations~(\ref{eq:multirho},~\ref{eq:multiz}) can be solved
numerically (for instance, using the Newton-Raphson method). The expectation value
of an observable $O$ at a reweighted $\beta$ can be expressed as
\begin{equation}
\langle O \rangle _{\beta} = \frac{\sum_{i = 1}^j \sum_{l = 1}^{N_i}
  g_i^{-1} O_i^l e^{- \beta E_i^l - f_\beta}} {\sum_{m=1}^J N_m
  g_m^{-1} e ^{- \beta_m E_i^l + f_m} } \ ,
\end{equation}
with 
\begin{equation}
e^{ - f_\beta \beta} = \frac{\sum_{i = 1}^j \sum_{l = 1}^{N_i}
  g_i^{-1}e^{- \beta E_i^a}} {\sum_{m=1}^J N_m
  g_m^{-1} e ^{- \beta_m E_i^a + f_m} } \ ,
\end{equation}
where all the $f_j$ (and $f_\beta$) are determined
self-consistently. In the previous two equations $E_i^l$ indicates the
value of the energy measured at Monte Carlo step $l$ in the run performed at
$\beta_i$ and likewise $O_i^l$ is the value of $O$ at Monte Carlo step
$l$ in the run at $\beta_i$. This method, introduced in~\cite{Ferrenberg}, is
known as {\em multi-histogram reweighting}. 

While multi-histogram reweighting has a wider range of
predictability and generally better precision than the
single-histogram method, there are still technical points to consider 
in order to apply the former technique efficiently. In particular, each
density of state value will receive contributions only from
simulations at which the corresponding energy is sampled with
sufficient accuracy. Notwithstanding this and other limitations, if carefully
implemented, multi-histogram reweighting is a powerful
tool for extracting to a very high degree of accuracy quantities related to
phase transitions such as critical exponents and critical couplings
from Monte Carlo simulations. While the obtained accuracy depends on the
details of the calculations (like the number of sampled $\beta$'s in the
critical region and the number of configurations generated at each
$\beta$, in addition to the chosen Monte Carlo update algorithm),
precisions well below the percent level on critical exponents and
significantly higher on critical couplings are within reach for a wide number of
statistical systems.    

\subsection{Bootstrap}
\label{sect:app:bootstrap}
The bootstrap technique is a general procedure that can be used to
obtain robust estimates of the standard error from observations of
variables even when the underlying probability distribution is
unknown. This technique proves particularly convenient when we are
interested in general functions of a random variable.

Let us assume we have a set of measurements of the variable
X, described by the ensamble $\{X_i\}$, $1 \le i \le N$, with $N$ the total number of
measurements. We are interested in the estimator of $F(X)$, where $F$
is a function of the variable $X$. The bootstrap provides the
estimator and the confidence interval according to the following
procedure: 
\begin{enumerate}
\item build the set of the estimators $\{ F_i = F(X_i)\} \ $;
\item for each $j$ with $1 \le j \le N_B$, with $N_B$ integer,
  construct a bootstrap sample by taking $N$ random values in 
  $\{F_i\}$ and call the resulting set $\{ F^j_i\}$ (each of this sets
  will be referred to as a bootstrap resample); 
\item for each bootstrap resample $\{F^j_i\}$, compute the average as
\begin{equation}
\overline{F}^j = \frac{1}{N} \sum_i F^j_i \ ; 
\end{equation}
\item an estimator for $F(X)$ is provided by the bootstrap average
\begin{eqnarray}
\bar{F} = \frac{1}{N_B} \sum_j\bar{F}^j \ , 
\end{eqnarray}
with the standard error given by 
\begin{eqnarray}
\Delta \bar{F} = \sqrt{\frac{1}{N_B - 1} \sum_j \left( \bar{F}^j -
  \bar{F}\right)^2 } \ .
\end{eqnarray}
\end{enumerate}
In practical applications, one takes $N_B$ of the order of 100-1000,
which ensures we fulfil the hypotheses of the central
limit theorem. 

Our discussion so far assumes lack of correlations between the data,
which is certainly not the case for Monte Carlo generated data. In order to
remove correlations from the sample, one applies a binning procedure, which consists in
computing averages over $N_b$ consecutive values of $F_i$ and replace
the latter subset of values with this average. This reduces the size
of the sample of the $F_i$ used in the bootstrap procedure from $N$ to $N/N_b$. If we
choose $N_b \gg \tau$, with $\tau$ the autocorrelation time, the data
in the reduced set are uncorrelated. We can then apply the bootstrap
procedure to the sample of the binned values. 

\section*{References}

\bibliography{svm-phtr}

\biboptions{sort&compress}

\end{document}